\PassOptionsToPackage{table}{xcolor}   
\documentclass[sigconf]{acmart}

\usepackage[utf8]{inputenc}
\usepackage{textcomp}
\usepackage{amsmath}

\usepackage{booktabs}
\usepackage{multirow}
\usepackage{enumitem}
\usepackage{makecell}
\usepackage{subcaption}
\usepackage{stfloats}
\usepackage{microtype}
\usepackage{graphicx}

\AtBeginDocument{%
  }

\copyrightyear{2026}
\acmYear{2026}
\setcopyright{cc}
\setcctype{by-nc-nd}
\acmConference[CHI '26]{Proceedings of the 2026 CHI Conference on Human Factors in Computing Systems}{April 13--17, 2026}{Barcelona, Spain}
\acmBooktitle{Proceedings of the 2026 CHI Conference on Human Factors in Computing Systems (CHI '26), April 13--17, 2026, Barcelona, Spain}
\acmPrice{}
\acmDOI{10.1145/3772318.3791747}
\acmISBN{979-8-4007-2278-3/2026/04}

\begin{document}

\title{Beyond Words: Measuring User Experience through Speech Analysis in Voice User Interfaces}

\author{Yong Ma}
\authornote{Both authors contributed equally to this research.}
\orcid{0000-0002-8398-4118}
\affiliation{%
  \institution{University of Bergen}
  \city{Bergen}
  \country{Norway}
}
\email{yong.ma@uib.no}

\author{Xuesong Zhang}
\authornotemark[1] 
\orcid{0000-0003-0722-4632}
\affiliation{%
  \institution{Southern University of Science and Technology}
  \city{Shenzhen}
  \country{China}
}
\email{hcisong@gmail.com}

\author{Xuedong Zhang}
\orcid{0009-0002-6361-1717}
\affiliation{%
  \institution{LMU Munich}
  \city{Munich}
  \country{Germany}
}
\email{xuedong.zhang@ifi.lmu.de}

\author{Natalia Bartłomiejczyk}
\orcid{0000-0002-1565-5494}
\affiliation{%
  \institution{Université de Neuchâtel}
  \city{Neuchâtel}
  \country{switzerland}
}
\email{natalia.bartlomiejczyk@unine.ch}

\author{Seungwoo Je}
\orcid{0000-0001-7418-4576}
\affiliation{%
  \institution{Southern University of Science and Technology}
  \city{Shenzhen}
  \country{China}
}
\email{seungwoo@sustech.edu.cn}

\author{Adrian Holzer}
\orcid{0000-0001-7946-1552}
\affiliation{%
  \institution{Université de Neuchâtel}
  \city{Neuchâtel}
  \country{switzerland}
}
\email{adrian.holzer@unine.ch}

\author{Morten Fjeld}
\orcid{0000-0002-9562-5147}
\affiliation{%
  \institution{University of Bergen}
  \city{Bergen}
  \country{Norway}
}
\affiliation{%
  \institution{Chalmers University of Technology}
  \city{Gothenburg}
  \country{Sweden}
}
\email{morten.fjeld@uib.no, fjeld@chalmers.se}

\author{Andreas Butz}
\orcid{0000-0002-9007-9888}
\affiliation{%
  \institution{LMU Munich}
  \city{Munich}
  \country{Germany}
}
\email{butz@ifi.lmu.de}

\renewcommand{\shortauthors}{Ma et al.}

\begin{abstract}
Voice assistants (VAs) are typically evaluated through task performance metrics and self-report questionnaires, but people’s voices themselves carry rich paralinguistic cues that reveal affect, effort, and interaction breakdowns. We present a within-subjects study (N=49) that systematically compared three VA personas across three usage scenarios to investigate whether speech-derived audio features can serve as a proxy for user experience (UX). Participants’ speech was analyzed for temporal, spectral, and linguistic markers, alongside standardized UX measures, brief mood and stress ratings, and a post-study questionnaire. We found correlations between specific speech features and self-reported satisfaction and experience. Furthermore, a machine learning model trained on speech features achieved promising accuracy in classifying UX levels, indicating that this might be a reasonable alternative to self-report instruments. Our findings establish speech as a viable, real-time signal for implicitly measuring UX and point toward adaptive VUIs that respond dynamically to emotional and usability-related vocal cues.
\end{abstract}



\maketitle

\section{Introduction}
Voice user interfaces (VUIs) are embedded into a range of everyday devices, from smart speakers and smartphones to in-car infotainment systems, supporting information seeking, control, and entertainment. Their adoption has enabled more natural, speech-based interaction, but it also raises questions about how users experience these systems when they perform well versus when they fail. Even though deployments scale, designers and researchers still rely predominantly on post-hoc questionnaires and coarse outcome measures (e.g., task success, completion time) to assess user experience (UX)~\cite{pugazhenthi2025quantitative,kirschthaler2020can}. While such measures are useful for after-the-fact evaluation, they miss the interaction’s in-the-moment dynamics. Because users need to reconstruct experiences from memory, fleeting episodes of frustration or relief are easily forgotten~\cite{pearl2016designing,nguyen2024enhancing}. Moreover, interviews or self-reports capture what participants say they felt and are vulnerable to recency bias, politeness effects, and limited self-awareness~\cite{snow2025breaking,klein2023design}; they can also interrupt interaction flow, heighten evaluation apprehension, and struggle to capture environmental and emotional context~\cite{hossain2017rethinking}. These biases can skew evaluation and, in turn, distort VUI design decisions, masking real user needs and weakening overall UX assessment.

To mitigate these biases, speech analysis offers a promising alternative for capturing UX indicators during VUI interactions. Beyond lexical content, user speech encodes a wealth of paralinguistic cues, including prosody (pitch, intensity, tempo)~\cite{scherer1973voice}, voice quality (jitter, shimmer, harmonic-to-noise ratio)~\cite{teixeira2013vocal}, disfluencies (filled pauses, self-repairs)~\cite{lasalle1995disfluency}, and social or interactional markers of engagement and activity~\cite{soman2010social,pentland2007social}, all of which convey critical affective information~\cite{el2011survey}.
These vocal properties have been consistently associated with psychological states such as affect, cognitive effort, interpersonal coordination, and uncertainty~\cite{kleinow2006potential, macpherson2019cognitive}, suggesting their potential as real-time, in-the-moment signals of UX during human-VUI interaction~\cite{fan2021older}.
Despite this promise, few evaluations capture in situ, turn-by-turn speech during actual VUI tasks; when audio is recorded, it is often limited to post-task interviews and analyzed alongside retrospective measures~\cite{ma2025measuring}. This approach severely limits the ability to link moment-to-moment acoustic cues to the user's perceived experience during the interaction. 
Consequently, there is limited guidance on which speech-derived features are most reliable, which UX constructs they represent (e.g., perceived attractiveness, trust, or naturalness), and whether these features can be effectively leveraged by contemporary AI/ML methods to automatically classify UX into different UX levels such as good, neutral, or bad.

We address this gap with a controlled, within-subjects study spanning three everyday scenarios that elicit functional, creative, and playful interaction styles, crossed with three assistant personas that operationalize design levers practitioners routinely tune (e.g., short vs. long response latency, clear vs. unclear voice, robust vs. brittle error handling). 
Our experimental prototype captures time-aligned audio, transcripts, and system events, including both perceived and internal latency, enabling clear separation between user behavior and system delays.
From each user utterance, we extract a compact yet interpretable set of features covering: prosody and timing (e.g., f0 level/variability, articulation rate, pause counts and durations), voice quality (jitter, shimmer, harmonic-to-noise ratio), spectro-temporal features, and transcript-based indicators (filled-pause and repair rates, command length). These features are paired with brief post-block ratings of \emph{attractiveness}, \emph{trust}, and \emph{satisfaction}, along with lightweight contextual covariates (mood, stress, self-reported talkativeness, audio quality).

By integrating synchronized multimodal data streams, including high-fidelity user speech, detailed system logs, and standardized UX measures, this work advances speech as a practical signal for evaluating UX in VUIs. We move beyond post-hoc questionnaires by demonstrating how compact, interpretable speech features systematically reflect interaction quality at the individual level. Specifically, we show that vocal cues correlate with key UX dimensions such as attractiveness, trust, and satisfaction, and that these cues can be used to distinguish positive, neutral, and negative UX using machine learning methods. Together, these contributions provide methodological guidance and empirical evidence for using speech as a real-time, low-friction proxy for UX evaluation and inform the design of adaptive VUIs that can respond dynamically to users’ vocal behavior during interaction.
Guided by these objectives, we investigate the following research questions:
\begin{itemize}[leftmargin=*]
\item \textbf{RQ1:} \textit{Can speech-derived features (e.g., pitch, speech rate, pause duration) serve as reliable indicators of UX?}
\item \textbf{RQ2:} \textit{How do these features correlate with attractiveness, trust and satisfaction?}
\item \textbf{RQ3:} \textit {Can these relevant speech features be used with AI/ML methods to identify different UX levels?}
\end{itemize}

Our contributions are threefold:
(1) a within-subjects, persona-manipulated VUI testbed and balanced protocol (3~personas $\times$ 3~scenarios, $N{=}49$) that align time-stamped audio, transcripts, and system-event traces with user-reported experience;
(2) an open, turn-level speech-feature pipeline suitable for real-time sensing and offline analysis; and
(3) empirical evidence and models showing that specific prosodic and temporal cues correlate with UX ratings and enable above-chance identification of UX levels, informing adaptive VUIs that respond to emotional and usability-related vocal cues as the conversation unfolds.

\section{Related Work}
\label{related_work}
Our work builds on related work in five different areas, which we briefly summarize below:
\subsection{Evaluating Voice User Interfaces}
The evaluation of VUIs has traditionally been anchored in two primary methodologies. The first involves objective, behavioral task metrics such as task success rate, completion time, number of errors, and turns-to-completion, which provide quantifiable measures of a system's efficiency and effectiveness~\cite{pugazhenthi2025quantitative,myers2018patterns}. The second, and perhaps more pervasive, approach relies on subjective self-report instruments administered after interaction. Standardized questionnaires like the System Usability Scale (SUS), AttrakDiff, and the User Experience Questionnaire (UEQ/UEQ+) are widely used to capture perceived usability, satisfaction, and hedonic qualities~\cite{schrepp2017construction,seaborn2021measuring,deshmukh2024user}. These methods are favored in both academic and industrial settings due to their scalability, ease of statistical comparison across studies, and straightforward administration through ubiquitous survey platforms~\cite{cheng2021can,demaeght2022survey,pal2019user}.

However, this reliance on post-hoc reflection is a fundamental limitation. These measures are inherently retrospective, making them susceptible to significant cognitive biases. Users must reconstruct their experience from memory, leading to recency effects where the final moments of an interaction disproportionately color the overall assessment, or to fading affect bias where the intensity of mid-task frustrations is forgotten~\cite{soleimani2017can,cockburn2017effects}. Furthermore, subjective measures are vulnerable to social desirability and politeness biases, where users may soften negative feedback when providing it directly to a researcher~\cite{hu2022polite,klein2023design}. Perhaps most critically, administering surveys or interviews inherently interrupts the natural flow of interaction, making users acutely aware they are being tested, which can alter their behavior and strain ecological validity~\cite{phukon2022can}.

To address these limitations, HCI researchers have turned to complementary approaches such as experience sampling methods (ESM)~\cite{faruk2024review,cherubini2009refined}, interaction telemetry and logging~\cite{naqvi2024code,paulik2021federated}, and physiological sensing, including heart rate variability~\cite{swoboda2022comparing} and EEG~\cite{gaspar2023measuring}. Nevertheless, each of these alternatives introduces its own trade-offs: experience sampling can be interruptive~\cite{mehrotra2015ask}, telemetry often lacks direct insight into affective states~\cite{soleimani2017can}, and physiological sensing imposes hardware requirements and raises privacy concerns~\cite{anandhan2025enhancing,kose2025beyond}.
These challenges motivate the use of speech as a low-friction, always-available signal that is inherently present in VUI interactions, offering a promising pathway to real-time, non-intrusive UX assessment.

\subsection{Speech as a Window into User State}
Research in psycholinguistics and human-computer interaction has established that prosodic features, such as fundamental frequency (F0), speech rate, and pausing behavior, can serve as reliable indicators of mental effort~\cite{kirkland2022s,munoz2020fundamental}, frustration~\cite{ang2002prosody, song2021frustration}, and engagement~\cite{yu2004detecting, ghosh2019automatic}. Similarly, voice quality measures including jitter, shimmer, and harmonic-to-noise ratio have been shown to reflect vocal tension and strain~\cite{van2018voice}. Disfluencies, such as filled pauses, restarts, and self-repairs, often signal cognitive planning difficulty or communicative misalignment~\cite{lickley2015fluency}. Moving beyond isolated features, research in social signal processing has demonstrated how combinations of speech cues can quantify higher-order constructs such as engagement, activity levels, vocal mirroring, and emphasis~\cite{curhan2007thin, pentland2010honest, soman2010social}. Within affective computing, such features have been used to infer emotional valence and arousal with considerable accuracy, particularly in controlled settings~\cite{d2018affective}.

For VUIs, these vocal characteristics offer a promising basis for real-time UX assessment, as they can be captured non-invasively using commodity microphones. Nevertheless, the use of speech as a sensing modality is not without challenges: vocal cues are highly sensitive to contextual factors such as speaker accent, room acoustics, and microphone quality~\cite{kishor2025voice}. Individual differences in vocal production further complicate cross-user generalization. Thus, robust speech-based UX sensing requires careful mitigation of confounds, such as through within-participant normalization, consistent audio recording conditions, and control of signal-to-noise ratio, to ensure that derived features genuinely reflect user state rather than extraneous variability.

\subsection{Speech-based UX Inference in HCI}
Beyond usability metrics in VUI evaluation~\cite{iniguez2021usability}, user satisfaction and engagement are central to UX assessment~\cite{pugazhenthi2025quantitative,alabbas2025weighted}, alongside constructs such as trust~\cite{klein2023design} and attractiveness~\cite{cordasco2014assessing}. HCI research increasingly links acoustic features to perceived satisfaction~\cite{egorow2017prediction,ma2025measuring}, trust~\cite{elkins2013sound,maltezou2025voice}, and attractiveness~\cite{wang2023pilot}, suggesting that speech analysis can provide unobtrusive indicators of these dimensions during the interaction. However, much of this evidence comes from post-hoc interview audio rather than in-situ, turn-level speech collected during tasks~\cite{ma2025measuring}, which establishes correlates but weakens ties to moment-to-moment experience and specific system behaviors. This gap motivates capturing consented, time-aligned speech during VUI interactions and aligning it with system events and concurrent UX judgments.

\subsection{System Behaviors as Design Levers}
The perceived performance of VUIs depends on more than automatic speech recognition (ASR) accuracy. System response latency, for instance, significantly influences perceived competence and responsiveness~\cite{bergmann2012second}. The design of confirmation behaviors and error-repair strategies contributes critically to transparency and user sense of control~\cite{karsenty2005transparency,nguyen2024enhancing}, while persona and conversational tone directly impact perceived warmth and likability~\cite{niculescu2013making,pias2024impact}. Previous research has demonstrated that extended delays can degrade trust and engagement, whereas timely confirmations and empathetic error handling help alleviate user frustration~\cite{li2025exploring,nguyen2024enhancing}.
However, many existing studies manipulate only one design factor at a time, employ highly scripted interactions, or rely exclusively on post-hoc surveys to evaluate UX outcomes~\cite{porcheron2018voice,dutsinma2022systematic}. This approach leaves a critical gap in understanding how these design choices manifest in users’ natural spoken behavior during real-time conversation. Uncovering this link is essential for developing VUIs capable of sensing interactional friction and adapting dynamically, for example, by simplifying responses when vocal cues indicate user fatigue, or increasing confirmation levels when disfluencies and pauses suggest uncertainty.

\subsection{Methods for Extracting Speech Features}
We compute two complementary families of acoustic features. Time-domain features (e.g., short-time energy, zero-crossing rate) describe amplitude and temporal structure directly from the waveform~\cite{jalil2013short,shrawankar2013techniques}. Frequency-domain features summarize spectral characteristics derived from short-time Fourier analysis, including Mel-Frequency Cepstral Coefficients (MFCCs), spectral centroid, bandwidth, roll-off, and related measures~\cite{tribolet2003frequency,le2011investigation}. We extract these features with librosa, a general-purpose Python audio analysis library~\cite{mcfee2015librosa}.
Following common practice, we also distinguish between low-level descriptors (LLDs) and high-level statistical functionals (HSFs)~\cite{xu2022multi}. LLDs are frame-wise measurements that capture fundamental acoustic properties (e.g., F0, energy, MFCCs), whereas HSFs aggregate LLDs over longer spans (e.g., per turn or block) using statistics such as means, standard deviations, percentiles, and deltas~\cite{xu2022multi}. To ensure coverage beyond purely spectral cues, we include prosodic and voice-quality measures, pitch/F0, jitter, shimmer, and harmonic-to-noise ratio -- computed with Praat/Parselmouth~\cite{jadoul2018introducing}. For standardized, reproducible sets (e.g., ComParE/eGeMAPS), we additionally rely on openSMILE~\cite{eyben2010opensmile}.
Together, these toolboxes yield a compact but expressive feature set spanning temporal, spectral, prosodic, and voice-quality dimensions suitable for both interpretable analyses and downstream modeling.

\subsection{Research Gaps}
Across these strands, two gaps persist: a need for turn-aligned, in-situ speech features recorded during real VUI tasks (not only in post-task interviews), and evidence clarifying which features reliably track specific UX constructs and whether small, interpretable sets can identify UX levels under subject-independent validation.
We address these gaps by (a) manipulating practical design levers (persona, latency, repair style) across varied scenarios, (b) logging consented, turn-level audio synchronized with system events, and (c) aligning speech features with standardized UX ratings. This allows mixed-effects analyses of how system behavior shapes vocal markers and tests whether compact feature sets can classify UX levels in ways useful for adaptive, real-time VUIs.
\section{Study Design}

\begin{figure*}[t]
\centering
\includegraphics[width=0.8\textwidth]{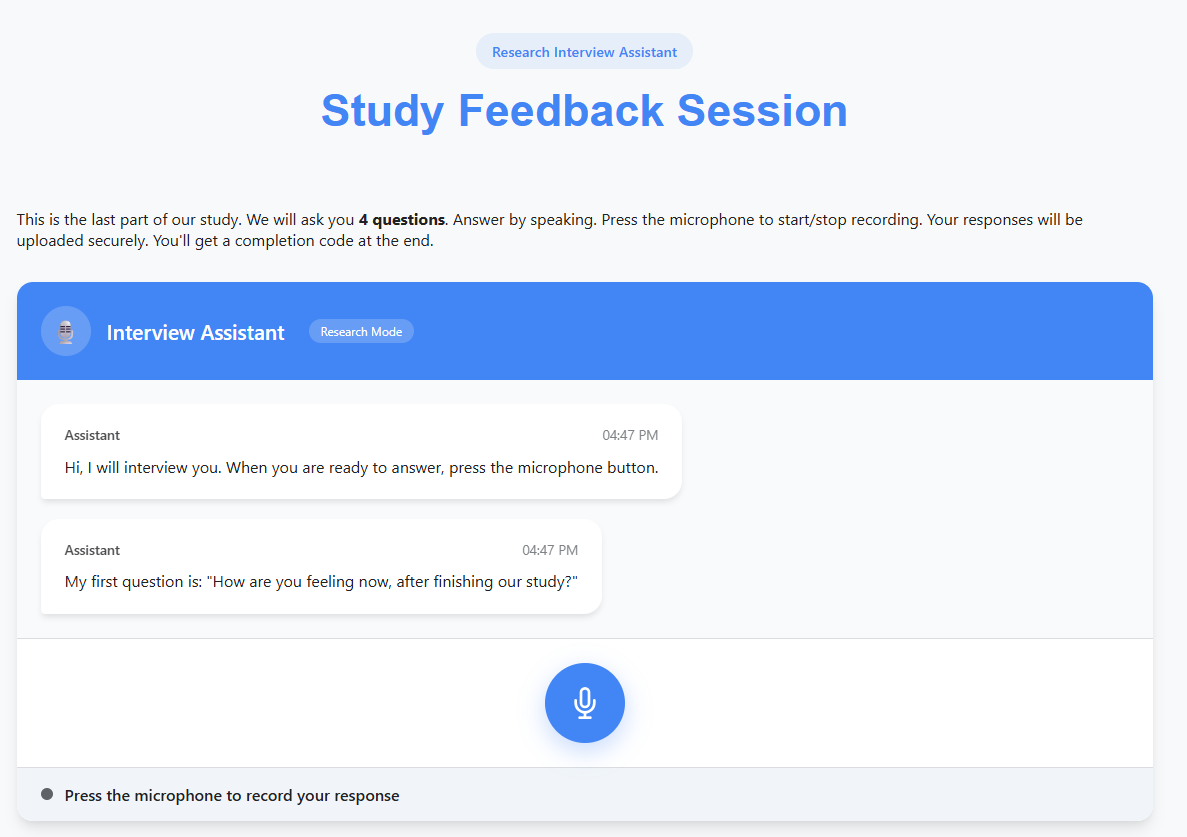}
\caption{The interface for the post-study interview, showing one of the four questions administered via the dedicated voice assistant (VA). This VA asked participants to reflect on and compare their experiences with the three experimental personas.}
\Description{Screenshot of the post-study interview interface showing a voice assistant prompt asking participants to reflect on and compare their experiences with three experimental voice assistant personas.}
\label{fig:interview_va}
\end{figure*}



To address our research questions, we conducted an online study on Prolific\footnote{\href{https://www.prolific.com/}{https://www.prolific.com/}} using a within-subject design. We created three distinct voice assistant (VA) personas by manipulating system factors such as response latency and error handling. 
To systematically evaluate how participants interacted with each VA persona, we defined three scenarios: functional, creative, playful.
This resulted in a 3 (Persona) $\times$ 3 (Scenario) within-subjects design, yielding nine unique experimental conditions.
Each participant interacted with all three VA personas across three different scenarios. This setup allowed us to capture variations in user experiences with the VA across various contexts.

\subsection{Overview of Experimental Procedure}

Our study procedure received approval from the ethics review board of SUSTech. We employed a 3 (Persona) $\times$ 3 (Scenario) full-factorial within-subjects design, yielding nine unique conditions. To mitigate order and sequence effects, we counterbalanced the assignment of personas to scenarios across participants using a Balanced Latin Square design~\cite{wagenaar1969note, mayer2020enhancing}. This ensured each persona-scenario combination was presented equally often and that potential learning or fatigue effects were systematically distributed.

The experimental procedure consisted of several phases. After reviewing an information sheet and providing digital informed consent, participants completed a pre-study questionnaire capturing demographics and baseline measures of mood and stress. Participants then engaged in three core interaction blocks. In each block, they were presented with a different VA persona-scenario pair, the sequence of which was determined by the counterbalancing scheme. Each interaction began with on-screen instructions for a specific scenario (e.g., trip planning), during which participants completed approximately five tasks via spoken dialogue. The system recorded all audio and logged interaction events with millisecond precision.

Immediately following each interaction, participants completed a post-block User Experience Questionnaire Plus (UEQ+) and reported their stress and mood. After engaging with all three VA personas, they were asked to complete a final comparative ranking, ordering the three VAs from best (1) to worst (3) and providing a brief open-ended rationale for their choices. Finally, participants took part in a structured interview with four predefined questions conducted by a dedicated interview VA (\autoref{fig:interview_va}) to gather richer qualitative feedback on their overall experience. The interview VA monitored for participant voice input and repeated questions when no response was detected to ensure all questions were addressed. After completing the interview, participants received a completion code to claim their compensation on Prolific. 
The interview data were analyzed using thematic analysis to identify recurring patterns in participants’ experiences and perceptions across the three VA personas.
The full details of the interview procedure and analysis are provided in Appendix~\ref{structured_interview}.


\subsection{Design of Voice Assistant Personas} 

Prior reviews~\cite{faruk2024review, 10.1145/3571884.3597129} have identified a broad set of UX dimensions relevant to VUIs, including system intelligence, response accuracy, response speed, and naturalness. Building on this work, we selected three key system factors, LLM features (output token length, personality, and error handling strategy), response latency, and voice quality, as our main manipulations. These factors correspond to established UX dimensions (e.g., efficiency, accuracy, intelligibility, and naturalness) and represent design levers that practitioners routinely adjust. By systematically varying them, we constructed three distinct VA personas, A, B, C, aimed at eliciting diverse user experiences (see \autoref{table1}). 
Additional example utterances illustrating the distinct conversational styles and error-handling behaviors of each persona are provided in Appendix~\ref{appendix:persona-examples}.

\begin{table*}[t]
\renewcommand{\arraystretch}{1.3}
\caption{Definition of VA persona A, B, and C: The design levers we varied to create the three VA Personas}
\Description{Table comparing three experimental VA personas across design dimensions including LLM output length, error handling behavior, personality style, response latency, and voice quality characteristics.}
\label{table1}
\centering
\begin{small}
\begin{tabular}{|c|c|c|c|}
\hline
\bfseries  & \bfseries VA persona A  & \bfseries VA persona B & \bfseries VA persona C\\
\hline
LLM Output Token Length & 100 & 300 & 600\\
Error Handling & acknowledges + repairs errors & no repair or clarification & silent or unhelpful\\
Personality & warm, competent & flat, emotionless & irritated, inconsistent \\
Response Latency & 0s & 4.5s & 9s\\
Voice Quality & clear & background noise & backgr. noise + added discontinuities \\ 
\hline
\end{tabular}
\end{small}
\end{table*}

\subsubsection{LLM and Response-Style Manipulation}
\label{LLM_VA}
We employed LLaMA-3.1-70B Instruct, accessed via the OpenRouter API,\footnote{\url{https://openrouter.ai/meta-llama/llama-3.1-70b-instruct}} as the backend language model for the VA. To experimentally manipulate perceived accuracy and conversational naturalness, we controlled two key generation parameters: the system prompt and the model’s maximum output token length. This approach is grounded in prior research establishing response verbosity as a significant factor in UX, particularly a known preference for concise replies in voice-based interactions \cite{kujala2014sentence,haas2022keep}.
We instantiated three distinct assistant personas, each defined by a unique prompt and token limit:
\begin{itemize}
\item \textbf{Persona A (100 tokens; Warm \& Competent):} Prompted to be helpful and empathetic. It paraphrases user requests, provides explicit confirmations, and proactively offers repairs for potential misunderstandings.
\item \textbf{Persona B (300 tokens; Neutral \& Terse):} Prompted to be strictly factual and minimal. It offers brief acknowledgments, avoids elaboration, and only performs repairs if explicitly instructed.
\item \textbf{Persona C (600 tokens; Irritated \& Inconsistent):} Prompted to be uncooperative and frustrated. It hedges, withholds or omits confirmations, and occasionally produces irrelevant or non-answer responses.
\end{itemize}

This design allowed us to vary communicative quality and verbosity systematically while holding core task content constant, thereby enabling a controlled investigation into how these stylistic features influence user UX.

\subsubsection{Response Latency}
Various studies~\cite{sogemeier2024importance, 10.1145/3334480.3382792, 10.1145/3409120.3410651} have demonstrated that response latency exerts a significant influence on user satisfaction and perceived usability. The findings converge on clear thresholds: latencies of under 2 seconds are generally considered acceptable; around 4 seconds, UX ratings drop substantially; and at 8 seconds or more, UX ratings are consistently rated at a very low level. These results highlight response latency as a critical determinant of UX and justify its inclusion as a primary manipulation factor in our study. In our design, Persona A was set to a 0s latency, meaning the LLM produced its response immediately after processing the input. Persona B had a 4.5s latency, where the LLM output was delayed accordingly, and Persona C featured a 9s latency.

\subsubsection{Voice Quality}
\label{TTS_Google}
We used the standard en-US-Neural2-J voice from the Google Text-to-Speech (TTS) API\footnote{\url{https://ai.google.dev/gemini-api/docs/speech-generation\#javascript}} as a high-quality baseline for all assistant personas. To experimentally manipulate perceived system quality and simulate real-world degradations, we programmatically altered the audio output for two personas (B and C) post-synthesis:
\begin{itemize}
\item \textbf{Persona B (Neutral \& Terse):} We introduced a layer of constant Gaussian white noise at a signal-to-noise ratio (SNR) of 15 dB to create a slightly degraded, low-fidelity listening experience suggestive of a low-quality hardware component or connection.
\item \textbf{Persona C (Irritated \& Inconsistent):} To signify an unstable and unreliable system, we added both the 15 dB noise from Persona B and simulated brief audio dropouts. This was achieved by randomly removing 100-300 ms segments of audio at intervals of 3-7 seconds, mimicking the effect of a severely impaired communication channel or processing errors.
\end{itemize}
 Persona A's audio remained unmodified, representing a clear and high-fidelity voice. This manipulation allowed us to carefully control the effect of voice quality on UX alongside the variations in LLM response style.

\subsection{Design of Scenarios}
\label{sec:scenarios}
We designed three scenarios to systematically evaluate the interaction of participants with each VA. 
Our aim was to create situations that combined everyday functionality with playful and imaginative use cases, ensuring both breadth of evaluation and sustained engagement. 
Each scenario consisted of five tasks and was limited to a maximum of ten minutes, allowing us to balance comparability across participants with sufficient depth of interaction. 
The tasks varied in complexity to elicit low, medium, and high levels of user engagement.

Our scenario design built on prior work by V\"{o}lkel et al.~\cite{voelkel2021va}, who developed a broad set of tasks to explore envisioned dialogues with VAs. 
While their task set included interactions involving Internet-of-Things (IoT) devices, we excluded such cases to ensure accessibility without additional hardware. 
Instead, we focused on tasks that can be accomplished with a standalone VA, such as information retrieval, reminders, or conversational play.  

The first scenario, \textit{Plan a Trip}, reflected common everyday applications of VAs, such as asking about the weather, requesting recommendations, or setting reminders. 
To structure these tasks into a coherent flow, we framed the participant’s role as that of a travel agent preparing a weekend trip to Rome for a client. 
This framing did not aim to simulate a professional setting but rather provided a controlled narrative that linked the individual tasks into a single storyline. 
By doing so, we avoided presenting participants with a series of unrelated requests, which would have led to short and fragmented exchanges. 
Instead, the travel agent context allowed us to design a scenario in which the tasks were thematically connected, while still being neutral enough to avoid personal questions or sensitive topics. 
This structure encouraged participants to engage in longer and more continuous interactions with the VA, thereby generating richer conversational data for subsequent analysis.  

The second scenario, \textit{Collaborative Storytelling}, was designed to elicit longer and more imaginative conversations. 
In this improvisation-based task, participants collaboratively co-created a story with the VA.
Participants initiated the narrative with a prompt and then iteratively contributed new ideas, while the VA expanded and adapted the storyline in response to each contribution.
This design was inspired by prior research on AI in theatre and improvisation, which has shown how performance practices can be used to explore social dynamics of human–agent interaction~\cite{luria2020roboticfut}. 
Recent systems such as ImprovMate further demonstrate how AI can act as a co-creator in improvisational contexts, supporting creativity and narrative coherence while embracing spontaneity~\cite{drago2025improvmate}. 
By framing the VA as a storytelling partner, we encouraged multi-turn, imaginative exchanges that went beyond functional queries.  

The third scenario, \textit{Fortune Teller}, introduced a role-playing element to evaluate the VA in a more performative and entertaining context. 
In this scenario, the assistant adopted the persona of a mystical oracle, and participants were invited to ask questions about their future across a range of life domains.
This scenario drew inspiration from work at the intersection of AI and live performance, particularly experiments in comedy and role-based improvisation, where AI has been used to sustain humorous or dramatic exchanges with human performers~\cite{mirowski2024robot}. 
The fortune-teller setup followed this tradition of playful role-based interaction, where dramatic and humorous elements pushed participants into longer, less predictable conversations.

\subsection{Platform for Generating the Experimental Conditions}

To investigate our research questions, we developed a web-based experimental platform that independently manipulates assistant persona and interaction scenario. The prototype was implemented as a single-page application using React and the Web Audio API, ensuring a consistent and responsive UX across all conditions.

\subsubsection{System Architecture and Persona Manipulation} For each condition, the prototype dynamically configured the backend to instantiate the target persona (A, B, or C). As detailed in Sections~\ref{LLM_VA} and~\ref{TTS_Google}, this involved:
(1) sending a corresponding system prompt and token limit to the LLaMA-3.1-70B model via the OpenRouter API, and
(2) applying specific audio degradations (added noise, dropouts) to the Google TTS output where applicable.
The interface provided task instructions, a visual context for the scenario (e.g., a story transcript in Fig.~\ref{fig:s2}), and clear visual feedback during recording and playback. The system handled full audio I/O, speech-to-text transcription, and dialogue management, logging all interaction events with high-precision timestamps for subsequent analysis.

This tightly controlled platform allowed us to isolate the effects of our manipulated variables (persona and scenario) on user speech and experience, forming the basis for our multimodal analysis.

\subsubsection{Scenario Presentation} 
\begin{figure*}[htb]
\centering
\begin{subfigure}[b]{0.32\linewidth}
\centering
\includegraphics[width=.9\linewidth]{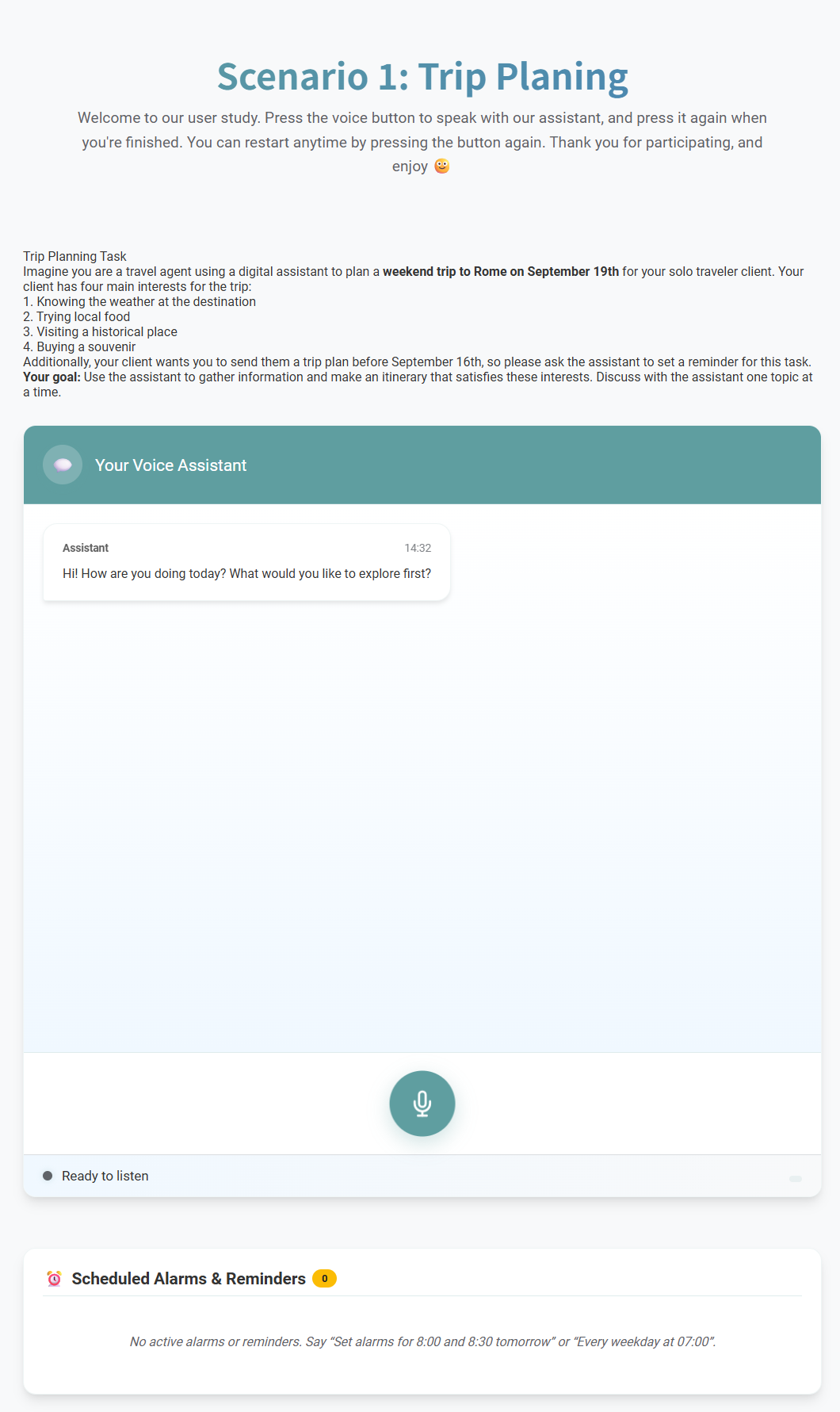}
\caption{Trip Planning: A goal-oriented task requiring precise information gathering.}
\label{fig:s1}
\end{subfigure}\hfill
\begin{subfigure}[b]{0.32\linewidth}
\centering
\includegraphics[width=.9\linewidth]{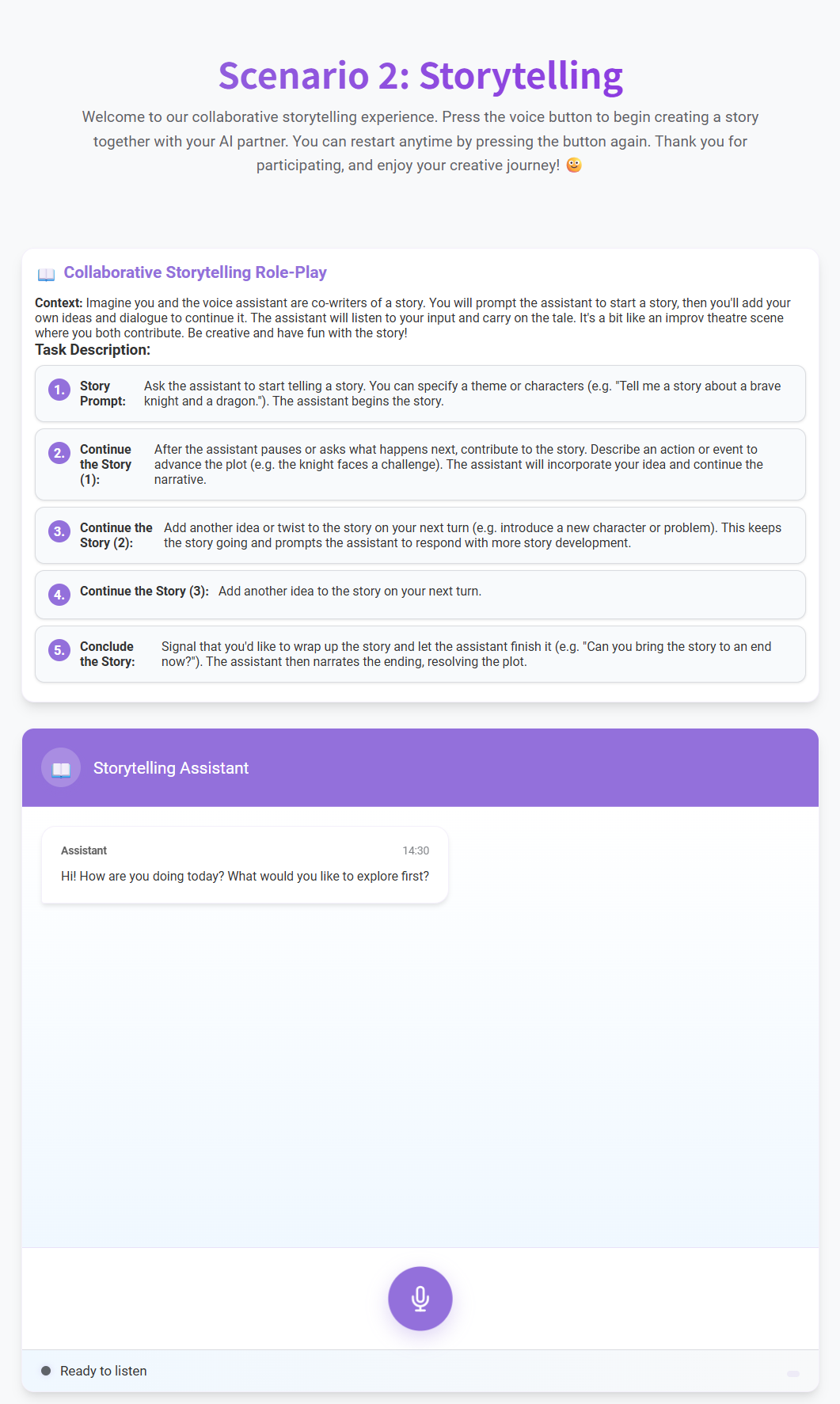}
\caption{Collaborative Storytelling: A creative, open-ended task requiring turn-taking.}
\label{fig:s2}
\end{subfigure}\hfill
\begin{subfigure}[b]{0.32\linewidth}
\centering
\includegraphics[width=.9\linewidth]{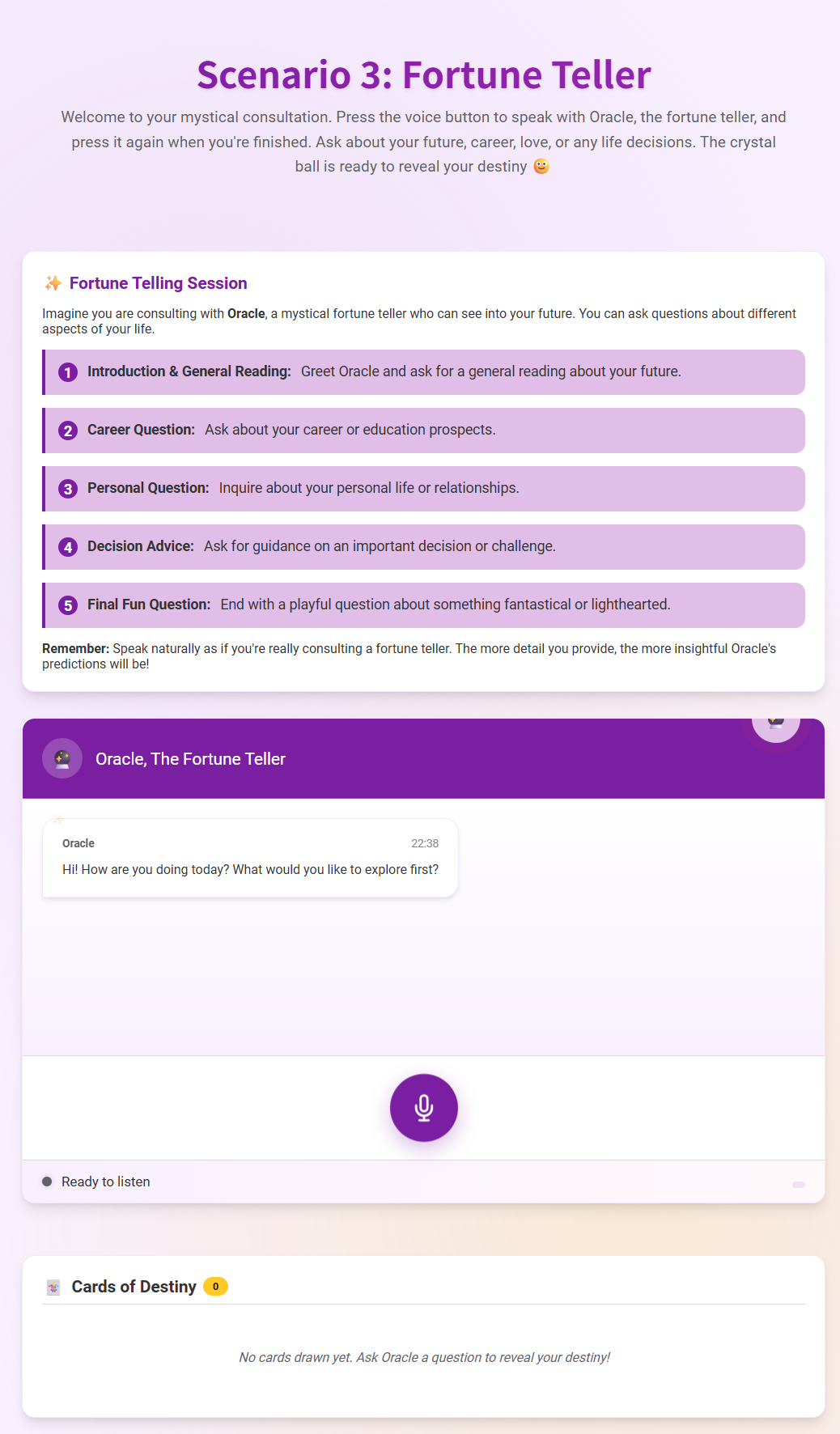}
\caption{Fortune Teller: A playful, socio-emo\-tional task with no ``correct'' outcome.}
\label{fig:s3}
\end{subfigure}
\caption{The three interaction scenarios as they were presented to participants, each designed to elicit different conversational patterns and user expectations: (a) a functional \textbf{Trip Planning} task, (b) a creative \textbf{Collaborative Storytelling} task, and (c) a playful \textbf{Fortune Teller} task.}
\Description{Three side-by-side screenshots showing the interaction scenarios used in the study: a trip planning interface focused on gathering travel details, a collaborative storytelling interface for creative turn-taking, and a fortune teller interface designed for playful socio-emotional interaction.}
\label{fig:scenarios}
\end{figure*}

The three distinct scenarios described in \autoref{sec:scenarios} were presented to participants as shown in \autoref{fig:scenarios}.

\section{Study Execution}

\subsection{Participants}

We recruited 52 participants through Prolific. 
After applying data-quality checks (e.g., for completed sessions and usable audio), we retained a final sample of 49 participants for analysis.
Eligibility was restricted to self-reported native English speakers residing in majority-English-speaking countries (e.g., the UK, US, Canada, Australia) to ensure language consistency.
All participants provided informed consent and were compensated £7.50 for the study, which had an estimated completion time of 50 minutes.  
More demographic details will be given in \autoref{sec:demographics} and in the Appendix \ref{tab:participants}.

\subsection{Apparatus}
The study was conducted remotely via a custom web-based prototype. The client-side application was built with React and the Web Audio API, which managed the user interface, dialogue flow, and audio recording (captured at 16kHz, 16-bit resolution) and playback. Audio files were saved in WAV format on our server for subsequent analysis.
The backend logic was powered by the LLaMA-3.1-70B Instruct model, accessed via the OpenRouter API. For speech synthesis, we used the Google Cloud Text-to-Speech API (voice: en-US-Neural2-J). The synthesized speech for each persona was programmatically degraded in post-processing to manipulate perceived voice quality, as detailed in Section~\ref{TTS_Google}.
Data processing and analysis were performed using a suite of specialized tools. Speech feature extraction was conducted using OpenSMILE\footnote{\url{https://www.audeering.com/research/opensmile/}} (for standard feature sets), Librosa\footnote{\url{https://librosa.org/doc/latest/feature.html}} (for spectral and temporal features), and Parselmouth\footnote{\url{https://sensein.group/senselab/senselab/audio/tasks/features_extraction.html}} (for prosodic and voice quality features like pitch, jitter, and shimmer). User speech recordings were automatically transcribed using the OpenAI Whisper\footnote{\url{https://openai.com/index/whisper/}} package. The resulting text data was then processed for linguistic analysis (e.g., disfluency counts, sentiment) using NLTK\footnote{\url{https://www.nltk.org/api/nltk.html}} and spaCy\footnote{\url{https://spacy.io/}}.
Finally, the Qualtrics\footnote{\url{https://www.qualtrics.com/}}  platform was used to develop and administer all questionnaires, and participants were recruited through the Prolific online platform.

\subsection{Measurements}
We employed a multi-faceted approach to capture both subjective UX and objective behavioral data. The primary subjective instrument was the UEQ+~\cite{klein2020construction, schrepp2021create, schrepp2019handbook}, administered after each interaction block to assess perceptions of the specific VA persona across 11 dimensions, including attractiveness, perspicuity, and dependability. To track affective fluctuations, we also collected self-reported mood and stress levels on 7-point Likert scales before the study and after each interaction. Following all three interactions, participants provided a comparative ranking of the three personas from best to worst, accompanied by a qualitative rationale. All questionnaire data was collected via Qualtrics.

A number of objective measures were derived from the recorded interactions. Participants' speech was transmitted as a Mono signal at 16~kHz sampling rate and 16~bit resolution for subsequent analysis. These settings provided sufficient audio quality and dynamic range for robust speech signal analysis, while reducing the necessary bandwidth to 18\% of that for CD-quality Audio. We also collected interaction logs with millisecond-precision timestamps, detailing system events such as turn-taking, response latency, and errors. These logs enabled the alignment of acoustic features with specific conversational contexts. Finally, structured qualitative feedback was gathered through an interview conducted by a dedicated interview VA (see \autoref{fig:interview_va}), capturing insights into participants’ overall experience.

\subsection{Speech Signal Analysis}
To analyze the vocal correlates of UX, we extracted a set of acoustic and linguistic features from the recorded speech data. All user utterances were processed using a suite of established computational tools. Low-level descriptors, including fundamental frequency (F0), energy, and spectral properties, were extracted using OpenSMILE~\cite{eyben2010opensmile}, implementing the standard ComParE feature set. To capture prosodic characteristics and voice quality, we utilized Parselmouth (a Python interface to Praat) to calculate metrics such as jitter, shimmer, and harmonics-to-noise ratio (HNR). Furthermore, Librosa was employed to compute additional spectral features like Mel-Frequency Cepstral Coefficients (MFCCs) and spectral contrast. Beyond acoustic features, we also performed linguistic analysis. Speech recordings were first transcribed using OpenAI's Whisper model for automatic speech recognition. The resulting text was then processed using NLTK and spaCy to quantify linguistic phenomena such as disfluency rates (e.g., filled pauses like ``um'' and ``uh'') and utterance length. This multi-tool approach yielded a rich, high-dimensional feature vector for each user turn, enabling us to investigate how specific vocal patterns shift with changes in perceived system quality and user state.

\section{Results From Questionnaire Analysis}
\label{questionnaire_label}
The questionnaires collected during the study served two purposes. First, they provided ground truth for correlating the speech analysis data and for training the ML model. Second, the results demonstrated that the design of the VA personas was effective in eliciting good, intermediate, and bad UX across multiple dimensions, such as mood and stress.

\subsection{Demographics}
\label{sec:demographics}
\begin{figure*}[t]
\centering
\includegraphics[width=\textwidth]{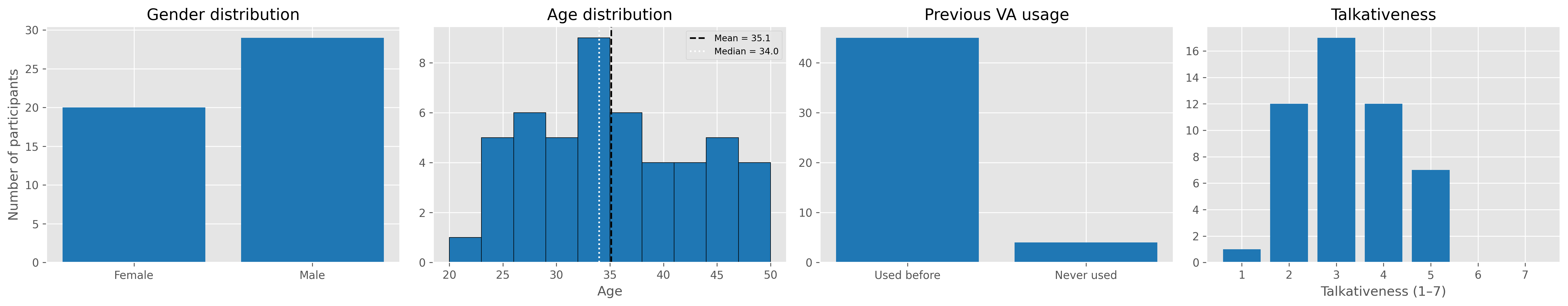}
\caption{Overview of participants gender, age and talkativeness.}
\Description{Bar charts summarizing participant demographics including gender distribution, age groups, and self-reported talkativeness levels.}
\label{fig:demographics}
\end{figure*}

Among the 49 participants who completed our study, the sample comprised 20 females and 29 males (see \autoref{fig:demographics}).
Participants’ ages ranged from 20 to 50 years, with a mean age of 35.1 years and a median age of 34 years. 
The majority of participants reported prior experience with VAs ($n = 45$), while only a small portion had never used one before ($n = 4$). 
Self-assessed talkativeness was distributed across the full 1--5 scale, with most participants rating themselves at the midpoints: 12 participants selected level 2, 17 selected level 3, 12 selected level 4, and 7 selected level 5, while only one participant reported being very quiet (level 1).
There were no participants who described themselves as more talkative (level 6 and 7).

\subsection{Mood and Stress}
A repeated-measures ANOVA revealed a significant main effect of \textit{VA} on mood ratings, $F(2, 96) = 25.65$, $p < .001$, indicating that participants’ mood significantly differed depending on which assistant they interacted with. 
There were also significant effects of scenario, $F(2, 96) = 3.30$, $p = .041$, and round, $F(2, 96) = 9.94$, $p < .001$ (see~\autoref{tab:anova_mood}).

\begin{table}[htbp]
\centering
\caption{Repeated-Measures ANOVA for Mood}
\Description{Table reporting repeated-measures ANOVA results for mood, including F values, numerator and denominator degrees of freedom, and p-values for the factors VA, Scenario, and Round.}
\label{tab:anova_mood}
\begin{tabular}{lcccc}
\toprule
\textbf{Factor} & \textbf{F Value} & \textbf{Num DF} & \textbf{Den DF} & \textbf{p-value} \\
\midrule
Voice Assistant & 25.65 & 2 & 96 & $<.001$ \\
Scenario        & 3.30  & 2 & 96 & .041 \\
Round           & 9.94  & 2 & 96 & $<.001$ \\
\bottomrule
\end{tabular}
\end{table}

\begin{table}[htbp]
\centering
\caption{Repeated-Measures ANOVA for Stress}
\Description{Table reporting repeated-measures ANOVA results for stress, including F values, numerator and denominator degrees of freedom, and p-values for the factors Voice Assistant, Scenario, and Round.}
\label{tab:anova_stress}
\renewcommand{\arraystretch}{1.2}
\begin{tabular}{lcccc}
\toprule
\textbf{Factor} & \textbf{F Value} & \textbf{Num DF} & \textbf{Den DF} & \textbf{p-value} \\
\midrule
Voice Assistant & 19.06 & 2 & 96 & $<.001$ \\
Scenario        & 2.32  & 2 & 96 & .104 \\
Round           & 3.39  & 2 & 96 & .038 \\
\bottomrule
\end{tabular}
\end{table}

Post-hoc comparisons using paired t-tests with Bonferroni correction (see~\autoref{tab:posthoc_mood}) showed that VA~A elicited significantly higher mood ratings ($M = 5.22$) than both VA~B ($M = 3.84$, $p < .001$) and VA~C ($M = 3.82$, $p < .001$). 
There was no significant difference between VA~B and VA~C ($p = 1.0$).
%
\begin{table*}[htbp]
\centering
\caption{Post-hoc Comparisons for Mood}
\Description{Table presenting post-hoc pairwise comparisons of mood ratings between VA personas A, B, and C, including mean values for each persona, t-statistics, and both uncorrected and Bonferroni-adjusted p-values.}
\label{tab:posthoc_mood}
\begin{tabular}{lcccccc}
\toprule
\textbf{Comparison} & \textbf{Mean A} & \textbf{Mean B} & \textbf{Mean C} & \textbf{t-statistic} & \textbf{p (uncorrected)}  & \textbf{p (Bonferroni)} \\
\midrule
A vs B & 5.22 & 3.84 & --     & 7.11 & $<.001$ & $<.001$ \\
A vs C & 5.22 & --     & 3.82 & 5.37 & $<.001$ & $<.001$ \\
B vs C & --     & 3.84 & 3.82 & 0.10 & .924   & 1.000 \\
\bottomrule
\end{tabular}
\end{table*}

\begin{figure}[htbp]
    \centering
    \includegraphics[width=0.45\textwidth]{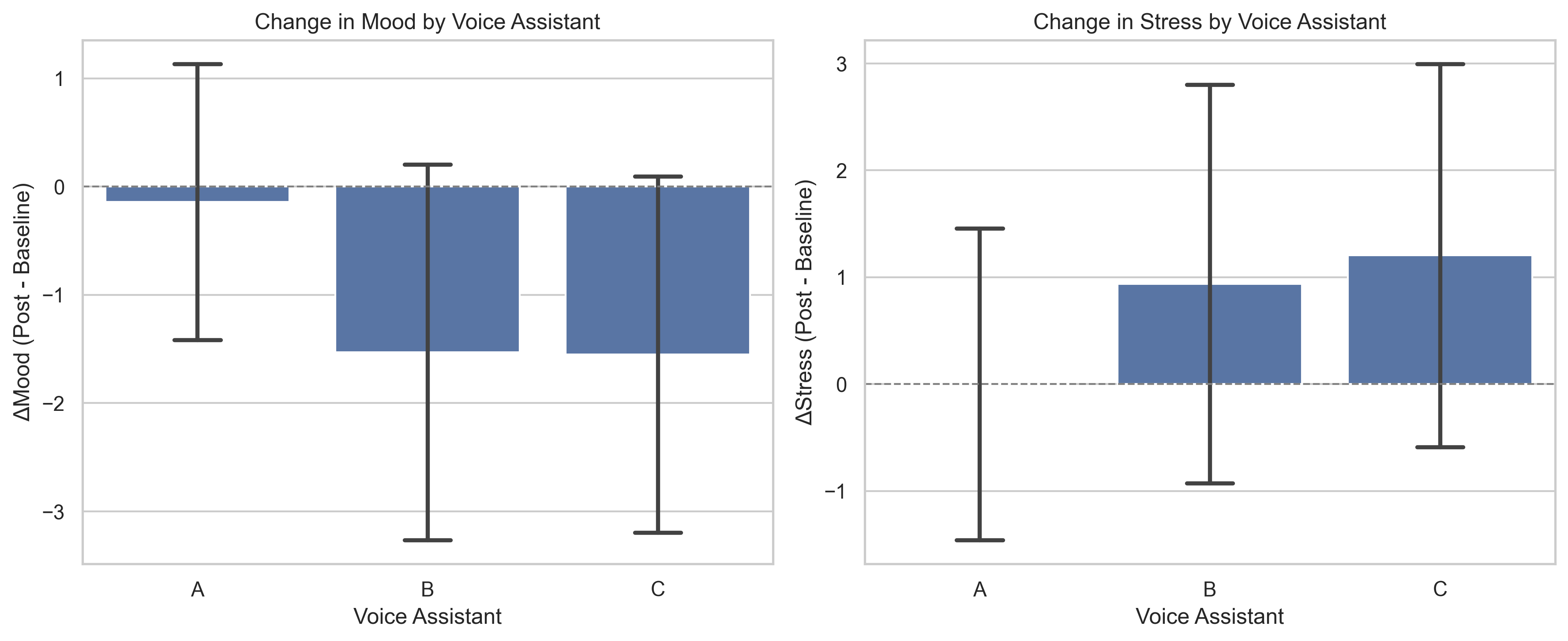}
    \caption{Mean change from baseline pre-study in Mood (left) and Stress (right) for each VA persona.}
    \Description{Combined plot showing mean changes from baseline in mood and stress levels across voice assistant personas, with mood changes displayed on the left and stress changes on the right.}
    \label{fig:delta_mood_stress}
\end{figure}

Baseline-adjusted values for Mood 
(see \autoref{fig:delta_mood_stress}) 
support this finding: mood remained nearly unchanged with VA~A, while it decreased after interaction with VA~B and VA~C.

\begin{figure}[htbp]
\centering
\begin{subfigure}[b]{0.48\linewidth}
    \centering
    \includegraphics[width=\linewidth]{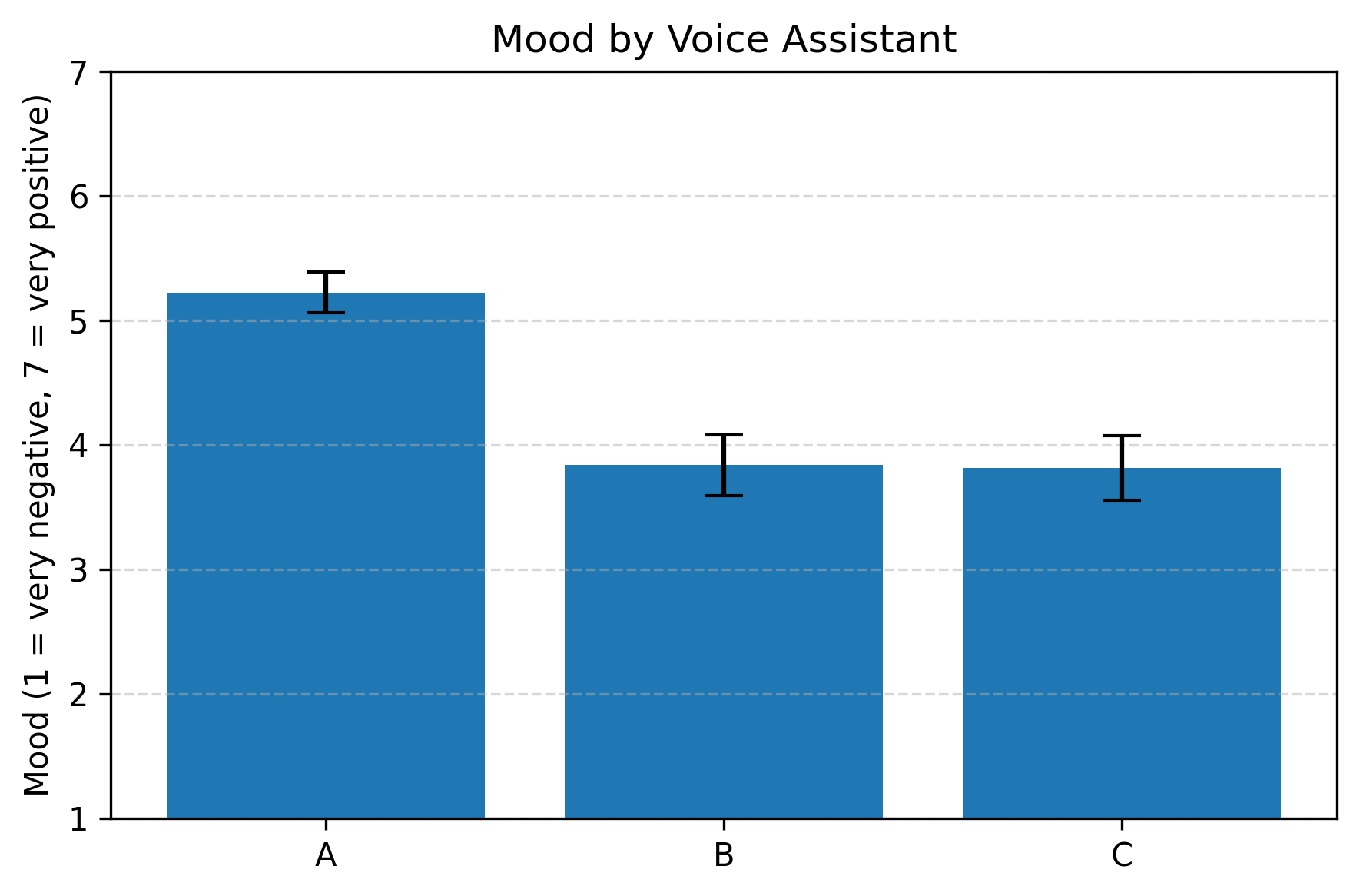}
    \label{fig:mood_va}
\end{subfigure}\hfill
\begin{subfigure}[b]{0.48\linewidth}
    \centering
    \includegraphics[width=\linewidth]{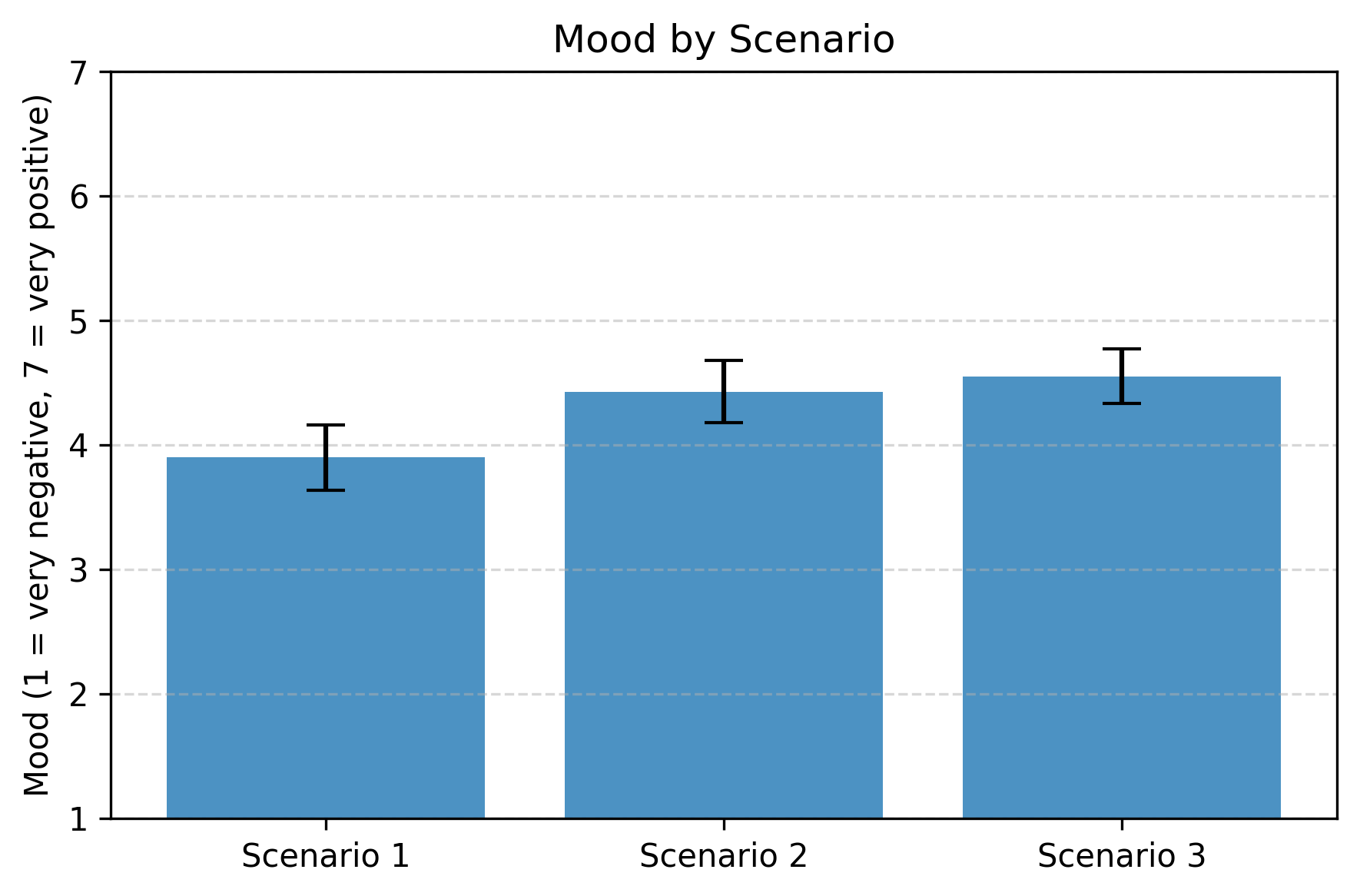}
    \label{fig:mood_scen}
\end{subfigure}
\caption{Mean mood ratings from all study rounds by VA (left) and Scenario (right).}
\Description{Two side-by-side plots showing average mood ratings across study rounds, grouped by voice assistant persona on the left and by interaction scenario on the right.}
\label{fig:mood_va_scenario}
\end{figure}

To illustrate overall patterns beyond change from baseline,~\autoref{fig:mood_va_scenario} shows the mean mood ratings for each VA (left) and Scenario (right). 
Mood was clearly higher with VA~A than the others, with smaller differences between scenarios.

For stress, the repeated-measures ANOVA showed a significant main effect of VA, $F(2, 96) = 19.06$, $p < .001$, and a smaller yet significant effect of round, $F(2, 96) = 3.39$, $p = .038$. 
The effect of scenario was not significant, $F(2, 96) = 2.32$, $p = .104$ (see~\autoref{tab:anova_stress}).


Post-hoc comparisons (see~\autoref{tab:posthoc_stress}) revealed that stress levels were significantly lower after interacting with VA~A ($M = 2.08$) compared to \textit{B} ($M = 3.02$, $p = .001$) and \textit{C} ($M = 3.29$, $p < .001$). 
No significant difference was observed between VA~B and VA~C ($p = .569$).

\begin{table*}[htbp]
\centering
\caption{Post-hoc Comparisons for Stress}
\label{tab:posthoc_stress}
\begin{tabular}{lcccccc}
\toprule
\textbf{Comparison} & \textbf{Mean A} & \textbf{Mean B} & \textbf{t-statistic} & \textbf{p (uncorrected)} & \textbf{Mean C} & \textbf{p (Bonferroni)} \\
\midrule
A vs B & 2.08 & 3.02 & -4.04 & .0002  & --     & .001 \\
A vs C & 2.08 & --     & -6.70 & $<.001$ & 3.29 & $<.001$ \\
B vs C & --     & 3.02 & -1.33 & .1897  & 3.29 & .569 \\
\bottomrule
\end{tabular}
\end{table*}

As shown in~\autoref{fig:delta_mood_stress}, Stress values increased most after using VA~C, followed by VA~B, while remaining lowest while using VA~A.
%

\begin{figure}[htbp]
\centering
\begin{subfigure}[b]{0.48\linewidth}
    \centering
    \includegraphics[width=\linewidth]{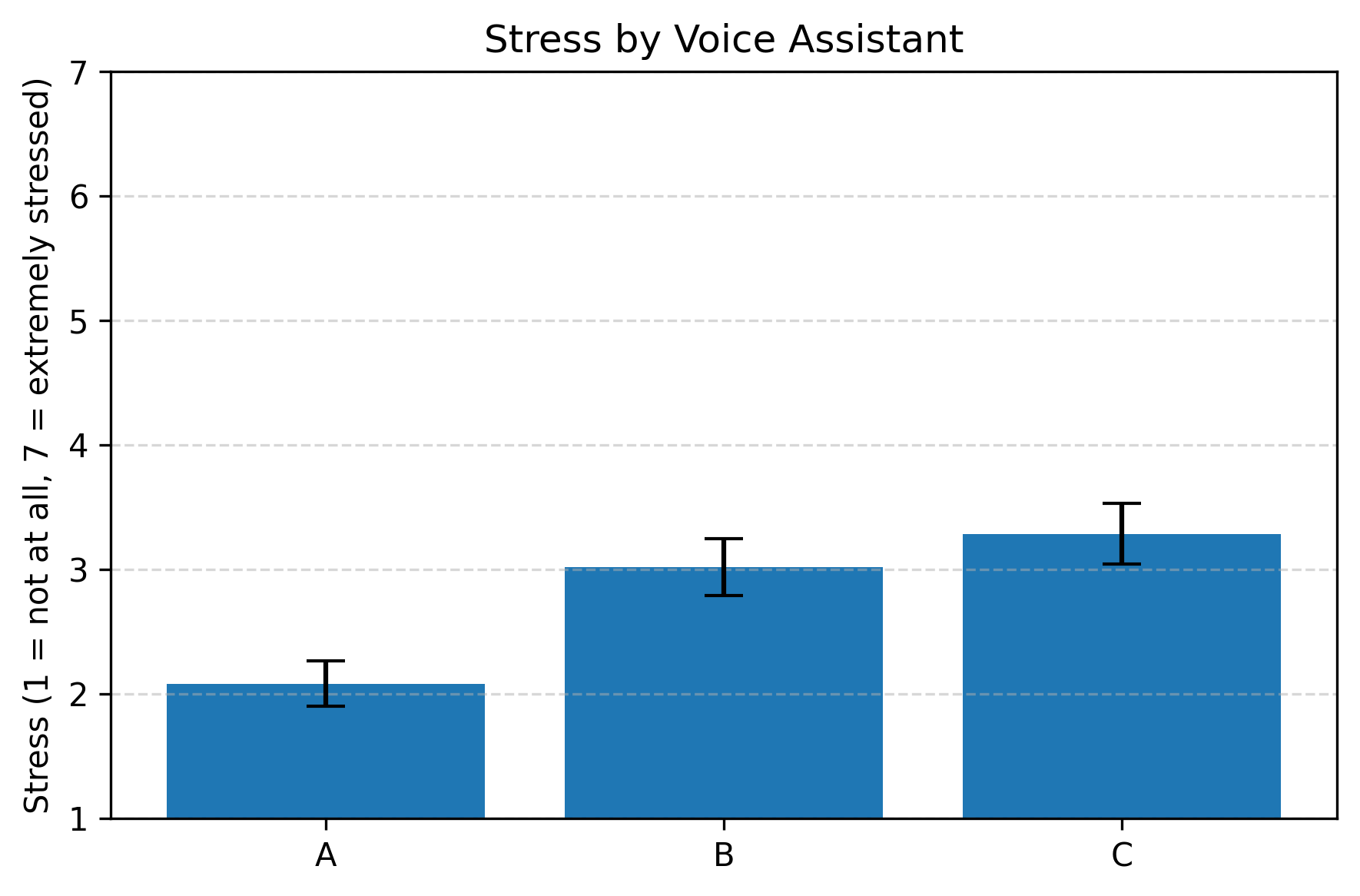}
    \label{fig:stress_va}
\end{subfigure}\hfill
\begin{subfigure}[b]{0.48\linewidth}
    \centering
    \includegraphics[width=\linewidth]{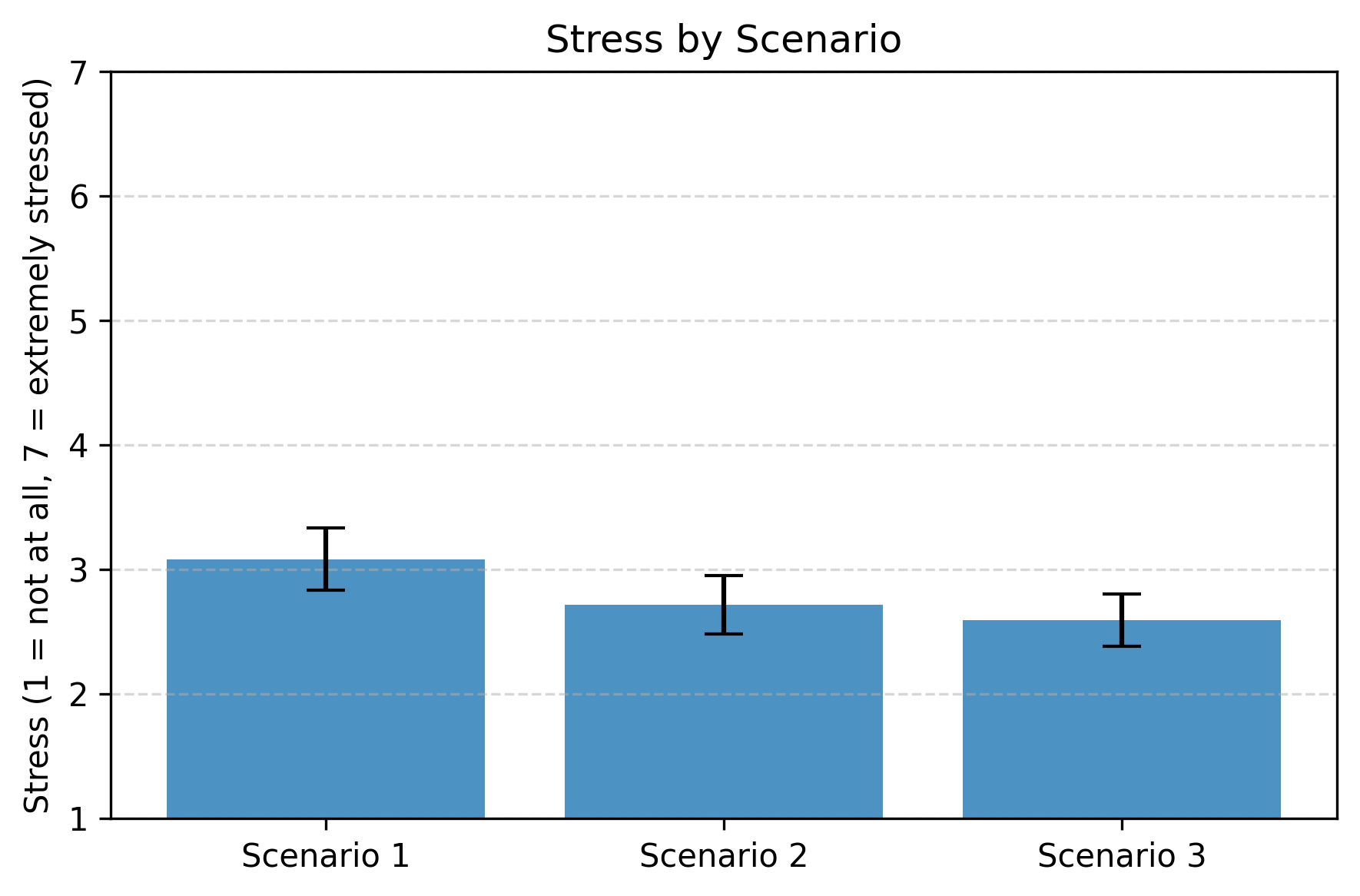}
    \label{fig:stress_scen}
\end{subfigure}
\caption{Mean stress ratings from all study rounds by VA persona (left) and Scenario (right).}
\Description{Two side-by-side plots showing average stress ratings across study rounds, with the left plot grouped by voice assistant persona and the right plot grouped by interaction scenario.}
\label{fig:stress_va_scenario}
\end{figure}

\autoref{fig:stress_va_scenario} illustrates the overall stress ratings across VAs and Scenarios. 
VA A resulted in lower average stress ratings, while stress appeared relatively consistent across scenarios.

\subsection{User Experience}
\label{VA_UX}
\begin{figure}[htbp]
    \centering
    \includegraphics[width=\linewidth]{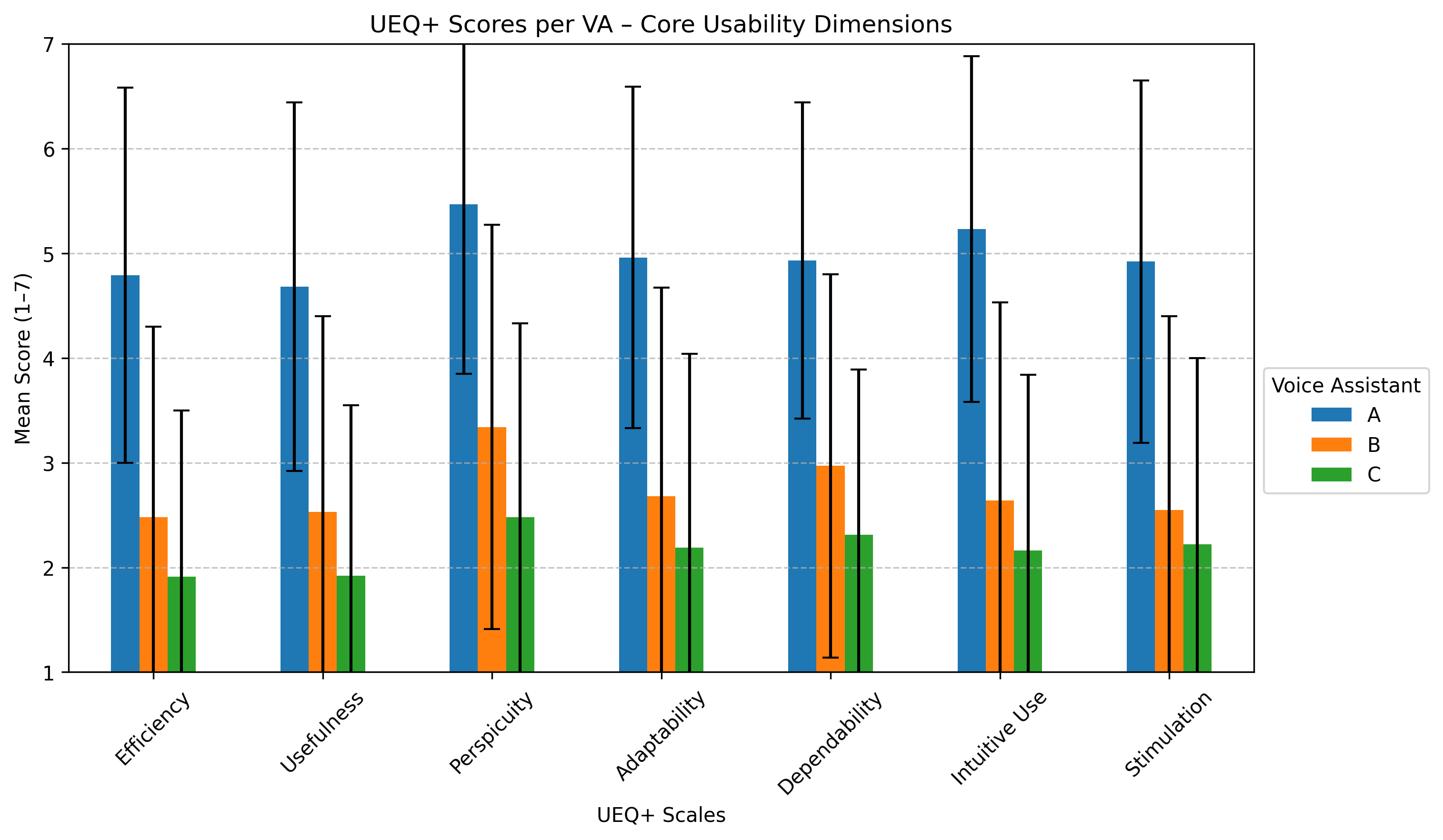}
    \caption{Mean UEQ+ scores for each VA on the core usability dimensions.}
    \Description{Bar chart showing mean UEQ+ usability scores for each voice assistant persona across core usability dimensions such as efficiency, dependability, clarity, and ease of use.}
    \label{fig:ueq1}
\end{figure}

Participants’ evaluations of the three VAs showed clear differences across usability and affective dimensions (see~\autoref{tab:ueq}). 
Overall, VA~A received consistently higher ratings than VA~B and VA~~C across nearly all UEQ+ scales.  
As shown on~\autoref{fig:ueq1}, VA~A achieved mean scores around or above 5 on a 7-point scale, including Efficiency ($M=4.79$), Usefulness ($M=4.68$), Perspicuity ($M=5.47$), Adaptability ($M=4.96$), Dependability ($M=4.93$), Intuitive Use ($M=5.23$), and Stimulation ($M=4.92$). 
In contrast, VA~B received moderate ratings across the same scales, with means between 2.48 (Efficiency)and 3.34 (Perspicuity), while VA~C was rated lowest, with scores generally between 1.91 (Effficency) and 2.48 (Perspicuity).  

\begin{figure}[h]
    \centering
    \includegraphics[width=\linewidth]{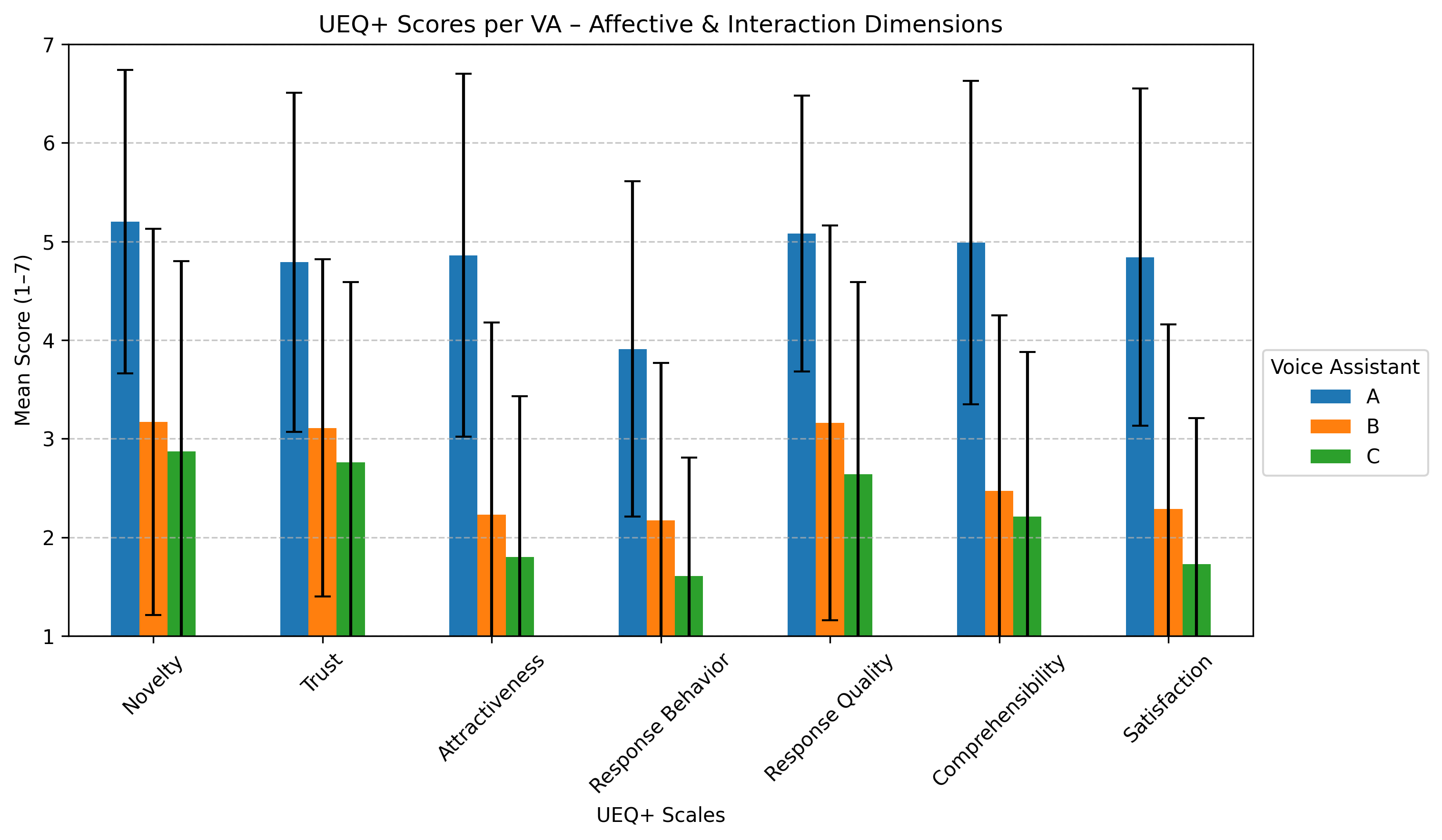}
    \caption{Mean UEQ+ scores for each VA on the affective and interaction dimensions.}
    \Description{Bar chart showing mean UEQ+ scores for each voice assistant persona across affective and interaction-related dimensions such as attractiveness, stimulation, and engagement.}
    \label{fig:ueq2}
\end{figure}

A similar pattern emerged is seen on~\autoref{fig:ueq2}. 
VA~A was again rated highest across Novelty ($M=5.20$), Trust ($M=4.79$), Attractiveness ($M=4.86$), Response Behavior ($M=3.91$), Response Quality ($M=5.08$), Comprehensibility ($M=4.99$), and Satisfaction ($M=4.84$). 
VA~B showed moderate evaluations, with scores ranging from 2.17 (Response Behavior) to 3.17 (Novelty). 
VA~C consistently received the lowest ratings, with most scales scoring between 1.61 (Response Behavior) and 2.89 (Novelty).  

These results indicate that VA~A was perceived as significantly more efficient, usable, and engaging compared to VA~B and VA~C, which both received lower and less consistent ratings.

\begin{table}[htbp]
\centering
\caption{Mean UEQ+ scores (1--7) for each VA across scales.}
\Description{Table listing mean UEQ+ usability and user experience scores for three voice assistant personas across multiple scales, including efficiency, usefulness, perspicuity, stimulation, trust, and satisfaction.}
\label{tab:ueq}
\renewcommand{\arraystretch}{1.15}
\begin{tabular}{lccc}
\toprule
\textbf{Scale} & \textbf{VA A} & \textbf{VA B} & \textbf{VA C} \\
\midrule
Efficiency         & 4.79 & 2.48 & 1.91 \\
Usefulness         & 4.68 & 2.53 & 1.92 \\
Perspicuity        & 5.47 & 3.34 & 2.48 \\
Adaptability       & 4.96 & 2.68 & 2.19 \\
Dependability      & 4.93 & 2.97 & 2.31 \\
Intuitive Use      & 5.23 & 2.64 & 2.16 \\
Stimulation        & 4.92 & 2.55 & 2.22 \\
Novelty            & 5.20 & 3.17 & 2.87 \\
Trust              & 4.79 & 3.11 & 2.76 \\
Attractiveness     & 4.86 & 2.23 & 1.80 \\
Response Behavior  & 3.91 & 2.17 & 1.61 \\
Response Quality   & 5.08 & 3.16 & 2.64 \\
Comprehensibility  & 4.99 & 2.47 & 2.21 \\
Satisfaction       & 4.84 & 2.29 & 1.73 \\
\bottomrule
\end{tabular}
\end{table}

\subsection{Feedback on Voice Assistants}
\label{VA_feedback}
\begin{figure}[htbp]
    \centering
    \includegraphics[width=.9\linewidth]{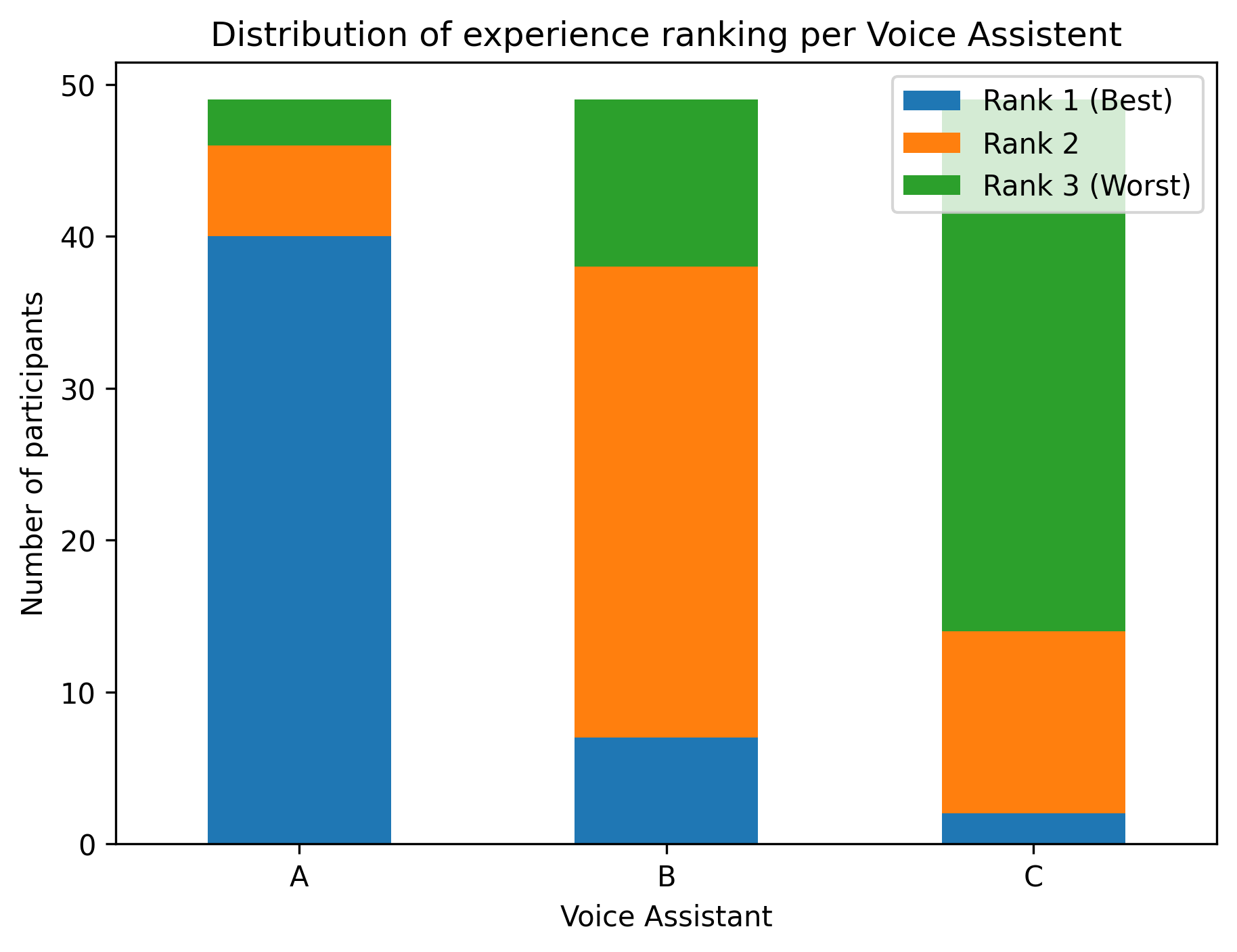}
    \caption{Distribution of experience rankings per VA.}
    \Description{Bar chart showing the distribution of participant experience rankings for each voice assistant persona, indicating how frequently each assistant was ranked highest, middle, or lowest.}
    \label{fig:ranking}
\end{figure}

In the survey, participants assessed not only usability of each VA.
They also ranked their experience with each VA from best to worst.
\autoref{fig:ranking} shows a clear pattern: 
VA~A was ranked as the best experience by the vast majority of participants ($n=40$), with only 6 people ranking it second and 3 ranking it last. 
This resulted in the lowest average ranking score ($M=1.24$), indicating a strong overall preference.  
By contrast, VA~B received mixed evaluations: it was ranked best by 7 participants, second by 31, and last by 11, leading to a moderate average ranking score ($M=2.08$). 
VA~C was least preferred overall, with only 2 participants ranking it first, 12 second, and 35 last. 
Its average ranking was the highest ($M=2.67$), reflecting its overall unfavorable perception.  
These quantitative rankings were further complemented by participants’ written explanations and their interview responses, which provided valuable insight into their reasoning and revealed what influenced their evaluations of VAs.

\subsubsection{Voice Assistant A}
Participants overwhelmingly praised VA~A, often citing its superior understanding and coherent responses across different scenarios. 
Many described VA~A as particularly helpful and natural in interaction. For example, one participant noted that \textit{``the trip planning VA~A was really helpful. It provided rational answers to all the requests and the voice feedback was also quite natural''} (P17). 
Others echoed that VA~A clearly understood user commands without errors or interference. 
Participant P6 wrote that VA~A \textit{``clearly understood my voice''} and had no background noise or interference, unlike the other two assistants. 
Such comments were typical for VA~A, whether it was used for trip planning or storytelling --users felt it matched their requests and responded accurately and fluently.

Despite this strong overall performance,  an outlier found VA~A less impressive as it struggled to stay on context. 
P47 complained that the fortune teller VA~A \textit{``spoke ok, but [I] seemed to be getting answers to a completely different subject''}. 
Such negative feedback on VA~A was rare, however. 
The vast majority of participants ranked VA~A as  their best experience, often emphasizing its reliability and understanding compared to the others.

\subsubsection{Voice Assistant B}
VA~B received mixed evaluations, with most participants placing it at an intermediate level.
Qualitative feedback suggests that while VA~B was not as beloved as VA~A, it typically performed better than VA~C. 
A common complaint was the presence of distracting background audio and slow or buggy responses. 
For instance, P1 found the trip planning VA~B nearly unusable due to \textit{``weird traffic background noise that wouldn’t stop,''} combined with the assistant ignoring commands and \textit{``repeating that it was receiving information''} without ever completing the task. 
Others encountered similar issues in the storytelling scenario, reporting that VA~B would repeat itself or misinterpret inputs. 
\textit{``The storytelling (VA~B) was a nightmare, it kept sending every message twice, repeating itself continuously and misunderstanding me regularly. It was very buggy and bad''}, wrote frustrated P24. 
Such glitches and lack of robustness were frequently cited as reasons VA~B could not be rated the best. Another even described the VA~B as \textit{``utter trash''} (P9) when it failed to provide relevant answers. 

However, a few participants had positive experiences with VA~B. 
In some scenarios, VA~B was able to deliver useful answers when the others failed.
Participant noted that VA~B at least responded with some correct information, it \textit{``at least gave me some of the information that I asked for''}(P41), even if it required repetition. 
These comments show that while VA~B had evident flaws, it was occasionally capable of a decent performance. 
In fact, a small number of participants (roughly 6 out of 52) did select VA~B as the best of the three, usually in cases where VA~A and VA~C had more serious failures. 
However, the consensus was that VA~B's background noise, repetition bugs, and comprehension problems kept it in second place behind VA~A in most participants’ rankings.

\subsubsection{Voice Assistant C}
VA~C was by far the least preferred assistant overall, with a large majority of participants describing their experience with it in negative terms. 
Many found VA~C effectively unusable due to technical failures. \textit{``The trip planning scenario (VA~C) was completely unusable and would not understand a single voice prompt''}, reported P5. 
Others had similar experiences of VA~C simply not responding or producing error messages instead of useful output. 
Several comments also pointed out that VA~C's voice delivery was unnaturally slow and garbled, which further undermined the experience. 
Participant P34, for example, explained that VA~C \textit{``spoke really slow, like a kind of stuttering, and didn't pronounce the words properly so some words didn’t even make any sense''}. 
This lack of fluency and responsiveness was a common refrain. 
It is not surprising, then, that 35 out of 49 participants ranked VA~C as their worst experience, reflecting an overall unfavorable perception of this assistant.
Only a handful of participants had anything positive to say about VA~C’s performance, and these tended to be isolated cases. 
A small minority (only 3 participants) actually ranked VA~C as the best, often because the other assistants failed even more dramatically in their sessions. 
For instance, P46 acknowledged that VA~C’s fortune-telling responses were \textit{``a bit confusing''} but still \textit{``clear and understandable''} in content.
However, these cases were exceptions. 
Nearly all other feedback on VA~C was resoundingly negative, emphasizing its inability to understand user requests, slow/glitchy audio, and tendency to produce errors or irrelevant answers. 
VA~C's poor performance was consistent across different scenarios (trip planning, storytelling, and fortune telling), making it the lowest-rated assistant by the vast majority of participants.


\subsection{UX Classification Based on UEQ+ KPI Benchmarks}
\label{ueq_kpi}
\begin{table*}[htbp]
\centering
\caption{UX levels (Positive, Neutral, Negative) and VA persona--scenario assignments (VA~A--S1 to VA~C--S3) across all 49 participants (P1--P49) over the three evaluation rounds (R1--R3).}
\Description{Large table showing each participant’s user experience level and corresponding voice assistant persona and scenario assignment across three evaluation rounds, illustrating the counterbalanced repeated-measures design.}
\label{tab:ux-va-clean}

\small
\setlength{\tabcolsep}{12pt}
\renewcommand{\arraystretch}{1.0}

\begin{tabular}{c c c c c c c}
\toprule
\multirow{2}{*}{Participant ID} &
\multicolumn{2}{c}{Round 1} &
\multicolumn{2}{c}{Round 2} &
\multicolumn{2}{c}{Round 3} \\
\cmidrule(lr){2-3}\cmidrule(lr){4-5}\cmidrule(lr){6-7}
& UX Level & VA Scenario & UX Level & VA Scenario & UX Level & VA Scenario \\
\midrule

P1  & Negative UX & VA C - S2 & Positive UX & VA A - S3 & Neutral UX & VA B - S1 \\
P2  & Positive UX & VA A - S3 & Neutral UX  & VA B - S1 & Negative UX & VA C - S2 \\
P3  & Neutral UX  & VA B - S2 & Negative UX & VA C - S3 & Positive UX & VA A - S1 \\
P4  & Positive UX & VA A - S1 & Neutral UX  & VA B - S2 & Negative UX & VA C - S3 \\
P5  & Positive UX & VA A - S2 & Neutral UX  & VA B - S3 & Negative UX & VA C - S1 \\
P6  & Positive UX & VA A - S1 & Neutral UX  & VA B - S2 & Negative UX & VA C - S3 \\
P7  & Neutral UX  & VA B - S1 & Negative UX & VA C - S2 & Positive UX & VA A - S3 \\
P8  & Positive UX & VA A - S3 & Neutral UX  & VA B - S1 & Negative UX & VA C - S2 \\
P9  & Negative UX & VA C - S2 & Negative UX & VA A - S3 & Negative UX & VA B - S1 \\
P10 & Positive UX & VA A - S2 & Neutral UX  & VA B - S3 & Negative UX & VA C - S1 \\
P11 & Positive UX & VA A - S2 & Neutral UX  & VA B - S3 & Negative UX & VA C - S1 \\
P12 & Positive UX & VA A - S3 & Neutral UX  & VA B - S1 & Negative UX & VA C - S2 \\
P13 & Negative UX & VA C - S3 & Positive UX & VA A - S1 & Neutral UX  & VA B - S2 \\
P14 & Neutral UX  & VA B - S3 & Negative UX & VA C - S1 & Positive UX & VA A - S2 \\
P15 & Neutral UX  & VA A - S3 & Negative UX & VA B - S1 & Negative UX & VA C - S2 \\
P16 & Negative UX & VA A - S3 & Negative UX & VA B - S1 & Negative UX & VA C - S2 \\
P17 & Positive UX & VA A - S1 & Neutral UX  & VA B - S2 & Negative UX & VA C - S3 \\
P18 & Negative UX & VA C - S3 & Positive UX & VA A - S1 & Neutral UX  & VA B - S2 \\
P19 & Positive UX & VA A - S2 & Neutral UX  & VA B - S3 & Negative UX & VA C - S1 \\
P20 & Neutral UX  & VA B - S2 & Negative UX & VA C - S3 & Positive UX & VA A - S1 \\
P21 & Neutral UX  & VA B - S1 & Negative UX & VA C - S2 & Positive UX & VA A - S3 \\
P22 & Negative UX & VA C - S2 & Positive UX & VA A - S3 & Neutral UX  & VA B - S1 \\
P23 & Negative UX & VA C - S1 & Positive UX & VA A - S2 & Neutral UX  & VA B - S3 \\
P24 & Neutral UX  & VA B - S2 & Negative UX & VA C - S3 & Positive UX & VA A - S1 \\
P25 & Neutral UX  & VA B - S3 & Negative UX & VA C - S1 & Positive UX & VA A - S2 \\
P26 & Negative UX & VA C - S3 & Positive UX & VA A - S1 & Neutral UX  & VA B - S2 \\
P27 & Neutral UX  & VA B - S1 & Negative UX & VA C - S2 & Positive UX & VA A - S3 \\
P28 & Negative UX & VA C - S2 & Negative UX & VA A - S3 & Negative UX & VA B - S1 \\
P29 & Negative UX & VA C - S2 & Positive UX & VA A - S3 & Neutral UX  & VA B - S1 \\
P30 & Neutral UX  & VA A - S3 & Neutral UX  & VA B - S1 & Negative UX & VA C - S2 \\
P31 & Positive UX & VA A - S1 & Neutral UX  & VA B - S2 & Negative UX & VA C - S3 \\
P32 & Negative UX & VA C - S1 & Neutral UX  & VA A - S2 & Negative UX & VA B - S3 \\
P33 & Neutral UX  & VA B - S2 & Negative UX & VA C - S3 & Positive UX & VA A - S1 \\
P34 & Negative UX & VA C - S1 & Positive UX & VA A - S2 & Neutral UX  & VA B - S3 \\
P35 & Positive UX & VA A - S2 & Positive UX & VA B - S3 & Negative UX & VA C - S1 \\
P36 & Positive UX & VA A - S1 & Positive UX & VA B - S2 & Positive UX & VA C - S3 \\
P37 & Positive UX & VA A - S3 & Neutral UX  & VA B - S1 & Negative UX & VA C - S2 \\
P38 & Neutral UX  & VA B - S3 & Neutral UX  & VA C - S1 & Neutral UX  & VA A - S2 \\
P39 & Negative UX & VA C - S2 & Positive UX & VA A - S3 & Neutral UX  & VA B - S1 \\
P40 & Positive UX & VA C - S3 & Positive UX & VA A - S1 & Positive UX & VA B - S2 \\
P41 & Negative UX & VA C - S2 & Negative UX & VA A - S3 & Negative UX & VA B - S1 \\
P42 & Positive UX & VA A - S2 & Neutral UX  & VA B - S3 & Negative UX & VA C - S1 \\
P43 & Negative UX & VA C - S3 & Positive UX & VA A - S1 & Neutral UX  & VA B - S2 \\
P44 & Neutral UX  & VA B - S3 & Negative UX & VA C - S1 & Positive UX & VA A - S2 \\
P45 & Negative UX & VA C - S3 & Positive UX & VA A - S1 & Neutral UX  & VA B - S2 \\
P46 & Positive UX & VA B - S2 & Positive UX & VA C - S3 & Positive UX & VA A - S1 \\
P47 & Neutral UX  & VA B - S1 & Negative UX & VA C - S2 & Positive UX & VA A - S3 \\
P48 & Neutral UX  & VA B - S2 & Neutral UX  & VA C - S3 & Positive UX & VA A - S1 \\
P49 & Negative UX & VA C - S1 & Positive UX & VA A - S2 & Neutral UX  & VA B - S3 \\

\bottomrule
\end{tabular}
\end{table*}

As presented in Section~\ref{VA_UX} and \ref{VA_feedback}, although our results show that, overall, VA~A is preferred, VA~B provides a moderate experience, and VA~C performs poorly, individual UX responses still vary considerably. 
As shown in Figure~\ref{fig:ranking}, even for the highest-rated VA~A, a subset of participants reported neutral or negative experiences.
To capture this individual-level variation, we categorized each participant’s UEQ+ scores across different VA personas and scenarios into positive, neutral, and negative UX. This approach can support future research in leveraging speech signals to identify individual-level shortcomings in otherwise high- and moderate-performing systems and improve them.

To categorize UX levels, we followed the interpretation guidelines proposed in the UEQ+ benchmark framework by Meiners et al.~\cite{meiners2024benchmark}. Their large-scale benchmark shows that mature commercial products typically achieve UEQ+ KPI values between +1 and +2, which reflects a clearly positive and competitive UX, whereas lower-performing products tend to cluster around KPI $\approx$ 0. Importantly, across the entire benchmark dataset, no evaluated product produced a negative KPI, indicating that values below zero represent UX substantially worse than established products. 

Based on this theoretical and empirical foundation, we defined three KPI-based UX categories that are generalizable, dataset-independent, and aligned with the UEQ+ interpretability framework:

\begin{itemize}
    \item \textbf{Positive UX:} KPI $\geq$ +1 \\
    (Comparable to strongly performing benchmark products)
    
    \item \textbf{Neutral UX:} $-1 \leq$ KPI $<$ +1 \\
    (Acceptable UX, but below the level of market-leading products)
    
    \item \textbf{Negative UX:} KPI $<$ $-1$ \\
    (Substantially below the expected UX standard of benchmarked products)
\end{itemize}

This classification avoids arbitrary, dataset-specific thresholds, such as percentile splits or unsupervised clustering, which limit cross-study comparability, and instead provides a theoretically grounded mapping of KPI values to UX levels that is portable across research contexts. This approach is consistent with prior UEQ+ work on KPI interpretation ~\cite{schrepp2019design, hinderks2019developing, kollmorgen2024selecting, santoso2022use, schrepp2019handbook}.

Table~\ref{tab:ux-va-clean} summarizes how these KPI-based UX levels were assigned to each participant (P1--P49) across the three evaluation rounds (R1--R3) and how they correspond to the specific VA personas (VA~A--C) and scenario conditions (S1--S3). This table provides the complete mapping used for all subsequent statistical analyses, including speech-feature comparisons across UX levels.

\section{Results From Speech Analysis}
\label{speech_analysis}

To investigate whether users’ speech reflects the quality of their interactions with three different VA Personas, we analyzed the recorded audio data across the three KPI-defined UX categories (Positive, Neutral, Negative). We organize the results according to our research questions: (1) differences in speech features across UX levels (RQ1), (2) associations between subjective UX dimensions and speech characteristics (RQ2), and (3) feasibility of predicting UX levels from speech using machine learning (RQ3).

\subsection{Speech Feature Variation Across Different UX Classification (RQ1)}
\label{ux_classification}

To examine whether speech acoustics systematically reflect perceived interaction quality, we compared speech features across the three benchmark-defined UX categories: \emph{Positive UX} ($\text{KPI} \geq +1$), \emph{Neutral UX} ($-1 \leq \text{KPI} < +1$), and \emph{Negative UX} ($\text{KPI} < -1$). These categories follow the UEQ+ KPI interpretation framework proposed by Meiners et al.~\cite{meiners2024benchmark} (see Section~\ref{ueq_kpi}) and the participant-level UX assignments summarized in Table~\ref{tab:ux-va-clean}. Applying this classification to our dataset yielded 48 audio recordings labeled as Positive UX, 45 as Neutral UX, and 54 as Negative UX. All speech recordings were grouped according to these UX labels for subsequent analysis.

We then extracted and compared a comprehensive set of vocal features spanning three key domains: (1) time-frequency acoustic features (e.g., spectral centroids, bandwidth, roll-off), (2) social speech features (e.g., speech activity, engagement rate), and (3) voice quality metrics (e.g., jitter, shimmer, HNR). 
This multi-dimensional feature set enabled us to characterize how users' vocal production systematically varied across different levels of perceived interaction quality.

Because the speech-feature distributions deviated from normality and the KPI-defined UX groups were substantially imbalanced in size, we employed a non-parametric statistical analysis pipeline. Overall group differences were evaluated using the Kruskal–Wallis H test~\cite{kruskal1952use,richardson2010nonparametric}, a rank-based omnibus method that is robust to skewed distributions and unequal variances. When the omnibus test was significant ($p < .05$), we performed post-hoc pairwise comparisons using Dunn’s test~\cite{dunn1964multiple}, applying a Bonferroni correction in line with recommended practice for multiple non-parametric comparisons~\cite{dinno2015nonparametric}. This analytical approach is widely employed in behavioral, speech, and HCI research when working with naturalistic acoustic data or uneven group structures~\cite{lee2024association,janssens2024child, zavichi2025gaze}. Applying this procedure provided a statistically rigorous basis for identifying which speech features reliably differentiate between Positive, Neutral, and Negative UX conditions.

\subsubsection{Results from Time- and Frequency-Domain Speech Features}

\begin{figure*}[t]
    \centering

    \begin{subfigure}[b]{0.32\linewidth}
        \centering
        \includegraphics[width=\linewidth]{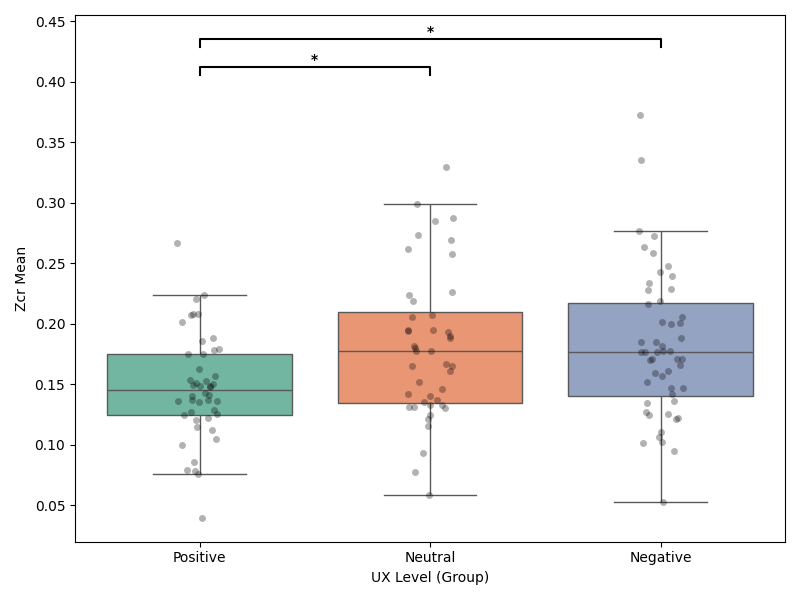}
        \caption{Zero-Crossing Rate (Mean)}
        \label{fig:zcr_mean}
    \end{subfigure}
    \hfill
    \begin{subfigure}[b]{0.32\linewidth}
        \centering
        \includegraphics[width=\linewidth]{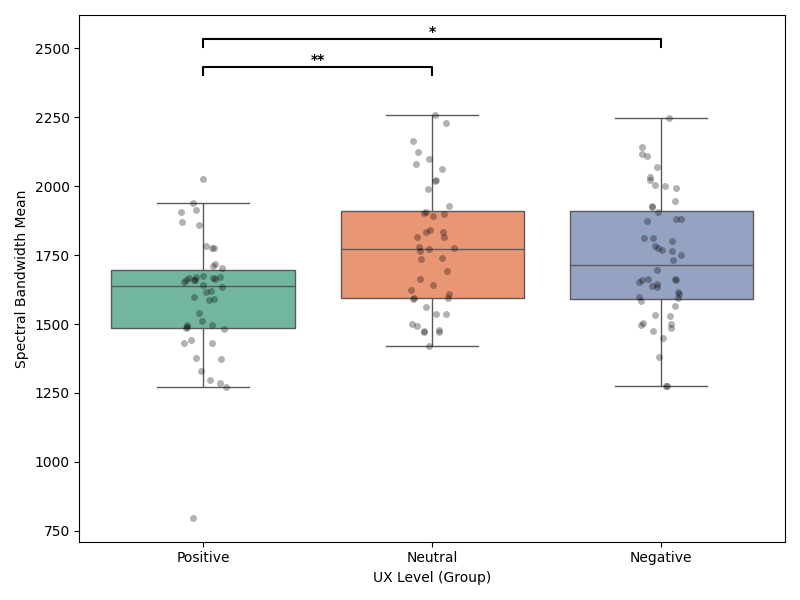}
        \caption{Spectral Bandwidth (Mean)}
        \label{fig:sbw_mean}
    \end{subfigure}
    \hfill
    \begin{subfigure}[b]{0.32\linewidth}
        \centering
        \includegraphics[width=\linewidth]{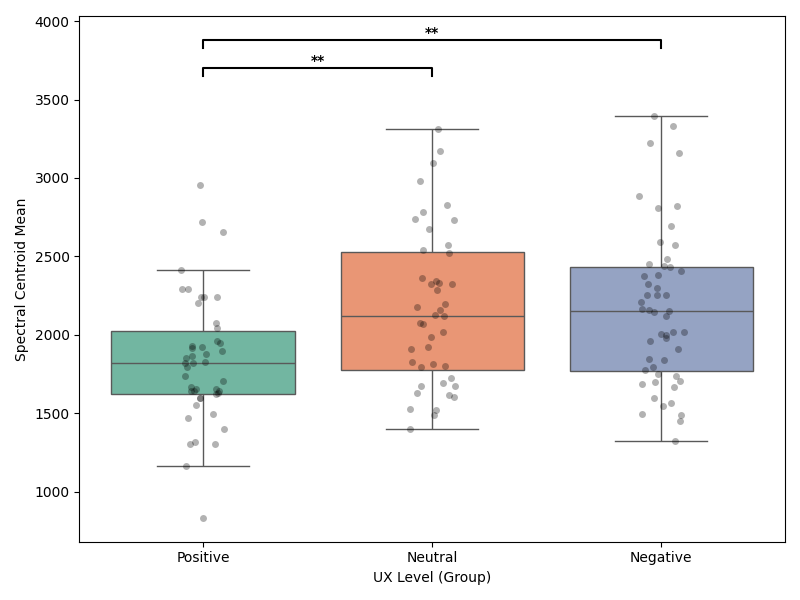}
        \caption{Spectral Centroid (Mean)}
        \label{fig:sc_mean}
    \end{subfigure}

    \caption{\textcolor{black}{Distributions of key time–frequency acoustic features across the three UX levels (Positive, Neutral, Negative). These plots show systematic increases in spectral dispersion (bandwidth), spectral brightness (centroid), and noisiness (ZCR) as UX quality decreases. Significance annotations indicate group differences based on the Kruskal–Wallis test with Dunn–Bonferroni post-hoc comparisons.}}
    \Description{Three side-by-side boxplots showing the distributions of zero-crossing rate, spectral bandwidth, and spectral centroid means across positive, neutral, and negative user experience levels, with statistical significance markers indicating group differences.}
    \label{fig:tf_three_features}
\end{figure*}

In this study, we extracted a comprehensive set of time- and frequency-domain speech features using the librosa and opensmile libraries in Python.  The extracted low-level descriptors included short-time energy, zero-crossing rate, fundamental frequency (pitch), spectral centroid, spectral bandwidth, and Mel-Frequency Cepstral Coefficients (MFCCs). 
To analyze the significance of feature differences across the three UX levels (Positive, Neutral, Negative), we conducted 
the Kruskal–Wallis test followed by Dunn’s post-hoc comparisons with a Bonferroni corrections.
The analysis revealed that several time–frequency descriptors differed reliably across Positive, Neutral, and Negative UX groups.
 Zero-Crossing Rate (ZCR$_{\text{mean}}$) exhibited a significant overall effect of UX level, $H=9.62$, $p=.008$, $\varepsilon^2=.055$, with Dunn--Bonferroni tests indicating that Negative UX interactions contained significantly higher ZCR than Positive UX ($p=.017$), suggesting increased noisiness and aperiodicity during less satisfactory interactions. Spectral Centroid Mean also showed a robust group effect ($H=13.71$, $p=.001$, $\varepsilon^2=.084$), with Negative UX speech being markedly brighter and richer in high-frequency energy than Positive UX ($p=.003$) and also significantly different from Neutral UX ($p=.005$). A similar pattern emerged for Spectral Bandwidth Mean, which varied significantly across groups ($H=11.51$, $p=.003$, $\varepsilon^2=.068$); post-hoc comparisons showed that Positive UX differed from both Neutral ($p=.0047$) and Negative UX ($p=.023$), with Negative UX interactions characterized by broader spectral dispersion.

Together, these features reveal a coherent acoustic signature of UX quality, such that as UX declines from Positive to Negative, speech becomes spectrally brighter, more dispersed, and noisier. Negative UX interactions consistently produced wider bandwidths, higher spectral centroids, and more zero-crossings, mirroring patterns associated with frustrated, strained, or effortful speech in prior speech-science research. Figure~\ref{fig:tf_three_features} illustrates these distributions for three representative features: ZCR, spectral centroid, and spectral bandwidth, highlighting clear separations between UX levels. These findings demonstrate that time–frequency speech features provide reliable, interpretable markers of UX and offer a statistically robust foundation for passive UX sensing in voice-based interfaces.

\subsubsection{Results from Social Speech Features}

\begin{figure*}[t]
  \centering

  \begin{subfigure}[b]{0.32\linewidth}
    \centering
    \includegraphics[width=\linewidth]{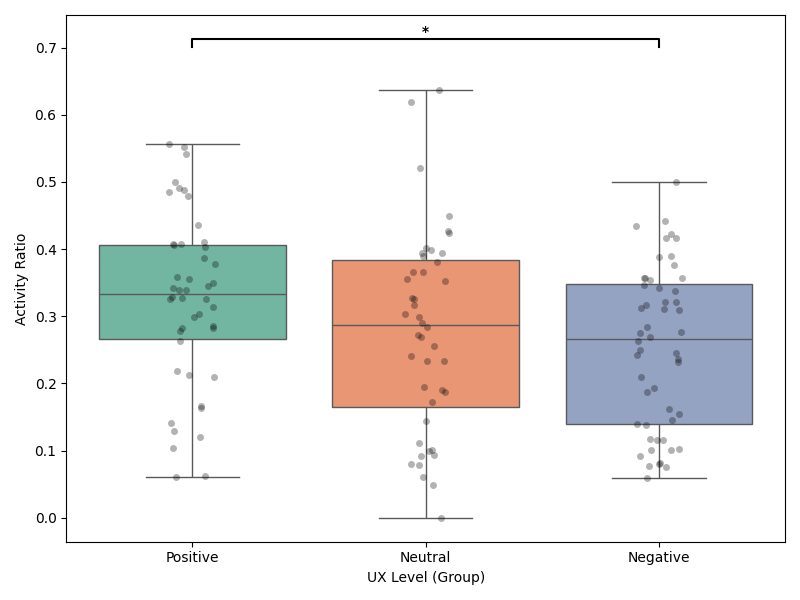}
    \caption{Speech Activity Ratio}
    \label{fig:social_activity_ratio}
  \end{subfigure}
  \hfill
  \begin{subfigure}[b]{0.32\linewidth}
    \centering
    \includegraphics[width=\linewidth]{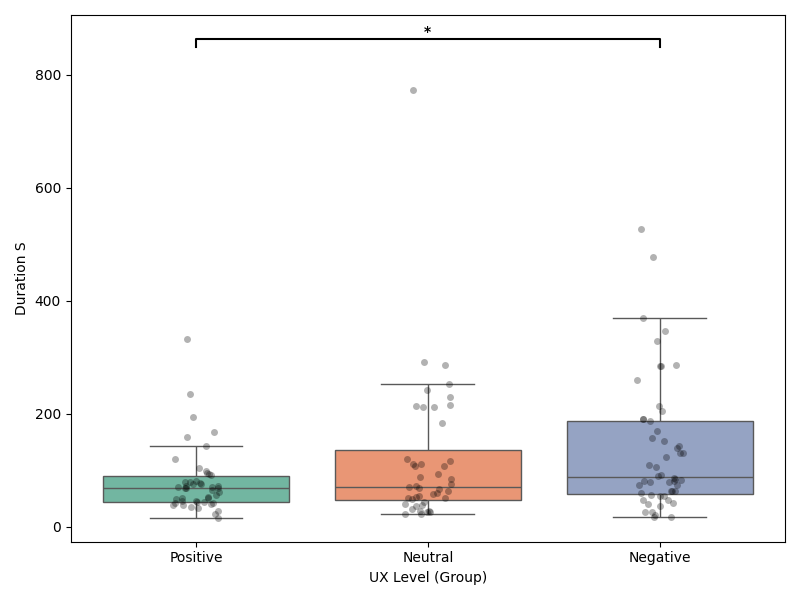}
    \caption{Interaction Duration (s)}
    \label{fig:social_duration}
  \end{subfigure}
  \hfill
  \begin{subfigure}[b]{0.32\linewidth}
    \centering
    \includegraphics[width=\linewidth]{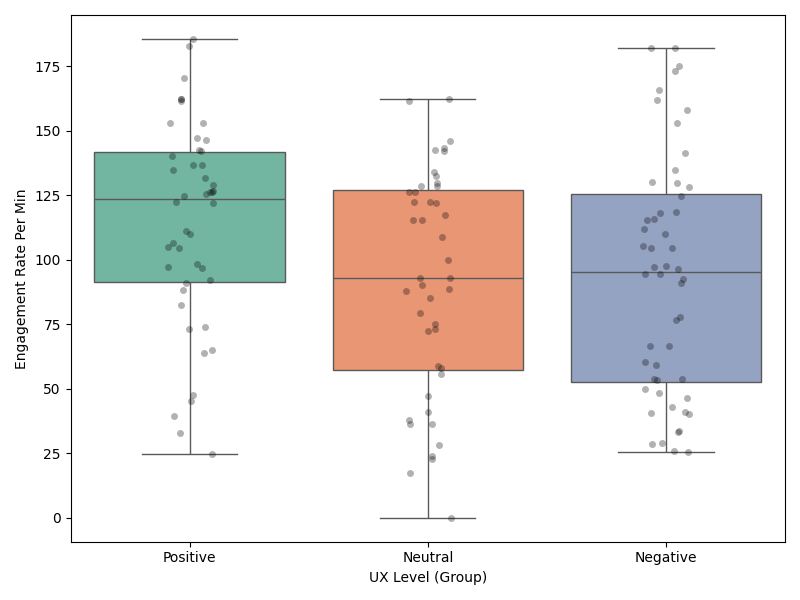}
    \caption{Engagement Rate (/min)}
    \label{fig:social_engagement}
  \end{subfigure}

  \caption{Distributions of social speech features across the three UX levels (Positive, Neutral, Negative). Boxplots show systematic differences in speech activity, interaction duration, and engagement rate as perceived UX quality changes; significance markers reflect Kruskal–Wallis tests with Dunn–Bonferroni post-hoc comparisons.}
  \Description{Three side-by-side boxplots illustrating speech activity ratio, interaction duration in seconds, and engagement rate per minute across positive, neutral, and negative user experience levels, with statistical significance indicators for group comparisons.}
  \label{fig:social_three_features}
\end{figure*}


To investigate how conversational behavior varied with perceived interaction quality, we analyzed all social speech features and the results showed that speech activity ratio, engagement rate per minute, and total speaking duration differed significantly across three UX levels. All features exhibited significant differences across the three UX groups based on Kruskal–Wallis tests, indicating that interaction quality systematically shaped how actively and fluently participants spoke during the VA tasks. 
Speech activity ratio showed a significant effect of UX level ($H = 7.14$, $p = .028$, $\varepsilon^2 = .037$), with Positive UX recordings containing more continuous speech than Negative UX recordings (Dunn–Bonferroni $p = .026$). As visualized in Figure~\ref{fig:social_activity_ratio}, participants contributed a larger share of the interaction when the experience was satisfying, whereas Negative UX conditions were marked by reduced vocal activity, suggesting hesitations, shorter turns, or disengagement. 
Speech duration (Interaction duration) also varied significantly across groups ($H = 6.67$, $p = .036$, $\varepsilon^2 = .034$). Negative UX interactions were substantially longer than Positive UX interactions (Dunn–Bonferroni $p = .030$), as shown in Figure~\ref{fig:social_duration}. Longer durations in the Negative UX condition likely reflect repeated clarifications, breakdowns, or extended repair sequences, a pattern consistent with conversational friction in prior HCI work.
Engagement rate per minute demonstrated a significant overall effect as well ($H = 7.33$, $p = .026$, $\varepsilon^2 = .038$). Although post-hoc comparisons approached significance for Positive vs. Neutral ($p = .055$) and Positive vs. Negative ($p = .057$), the trend (Figure~\ref{fig:social_engagement}) shows higher interaction density under Positive UX conditions and a notable reduction in both Neutral and Negative UX. 
This suggests that fluid, rapid turn-taking, indicative of confident and smooth task progression, declines when users experience friction or frustration.

Overall, these results reveal a coherent behavioral signature: as UX quality worsens, users speak less continuously, interact more slowly, and require longer interactions to complete the same tasks. Positive UX interactions were characterized by higher activity, faster engagement, and shorter task duration, whereas Negative UX interactions were marked by reduced vocal contribution and prolonged, effortful exchanges. These findings demonstrate that social speech patterns provide robust, behaviorally grounded indicators of moment-to-moment UX quality in voice-based interactions.

\subsubsection{Results from Voice Quality Metrics}

\begin{figure*}[t]
    \centering

    \begin{subfigure}[b]{0.32\linewidth}
        \centering
        \includegraphics[width=\linewidth]{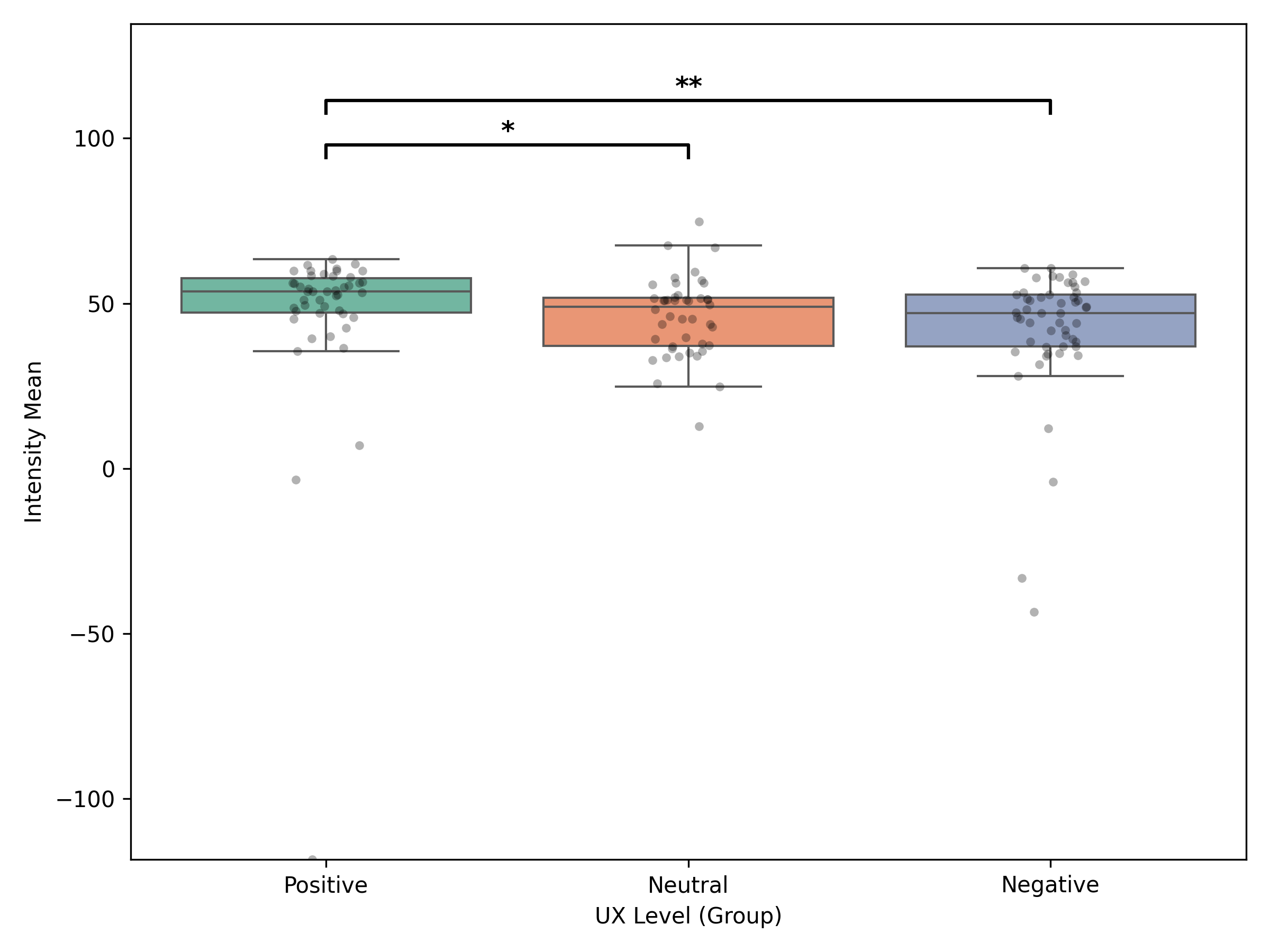}
        \caption{Intensity (Mean)}
        \label{fig:intensity_mean}
    \end{subfigure}
    \hfill
    \begin{subfigure}[b]{0.32\linewidth}
        \centering
        \includegraphics[width=\linewidth]{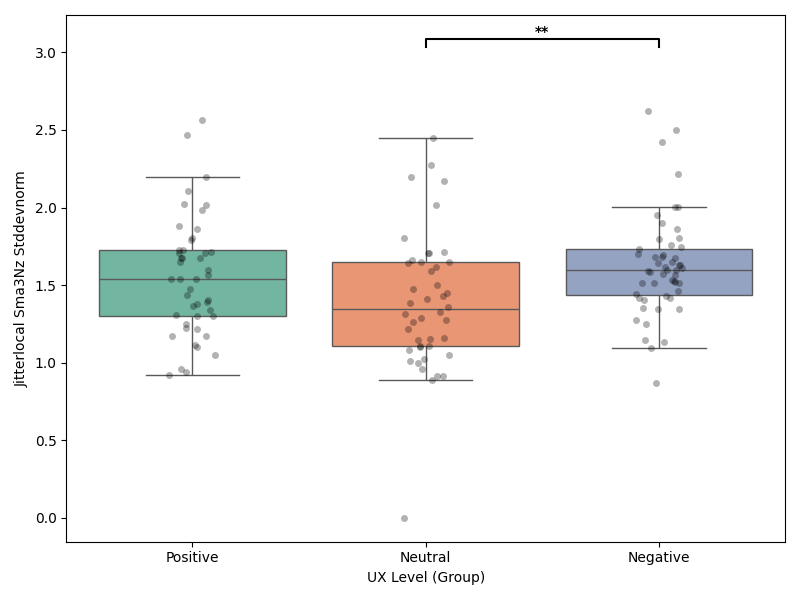}
        \caption{Jitter (std)}
        \label{fig:jitter_std}
    \end{subfigure}
    \hfill
    \begin{subfigure}[b]{0.32\linewidth}
        \centering
        \includegraphics[width=\linewidth]{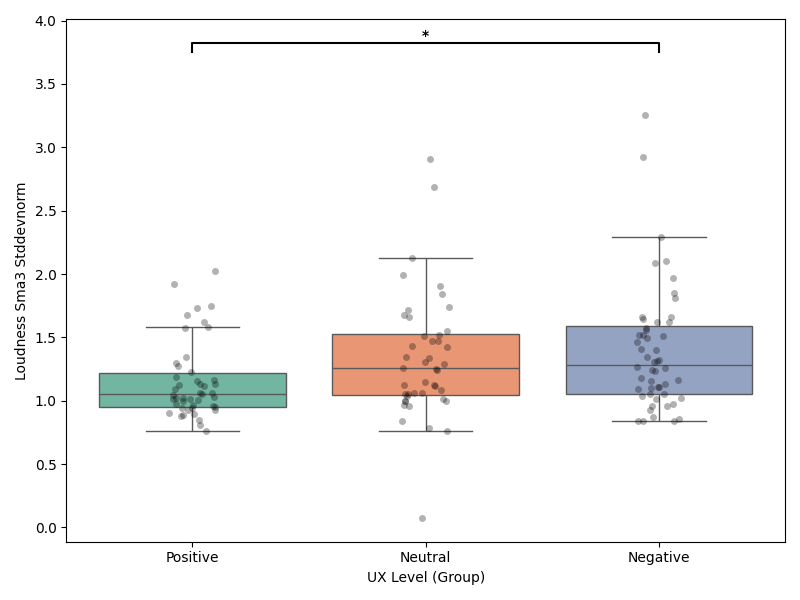}
        \caption{Loudness (std)}
        \label{fig:loudness_std}
    \end{subfigure}

    \vspace{0.8em}

    \begin{subfigure}[b]{0.32\linewidth}
        \centering
        \includegraphics[width=\linewidth]{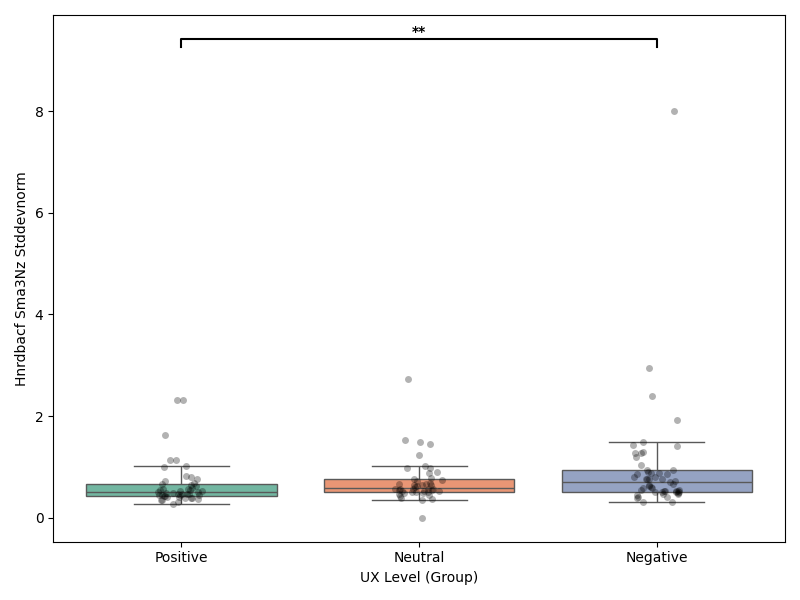}
        \caption{HNR (std)}
        \label{fig:hnr_std}
    \end{subfigure}
    \hfill
    \begin{subfigure}[b]{0.32\linewidth}
        \centering
        \includegraphics[width=\linewidth]{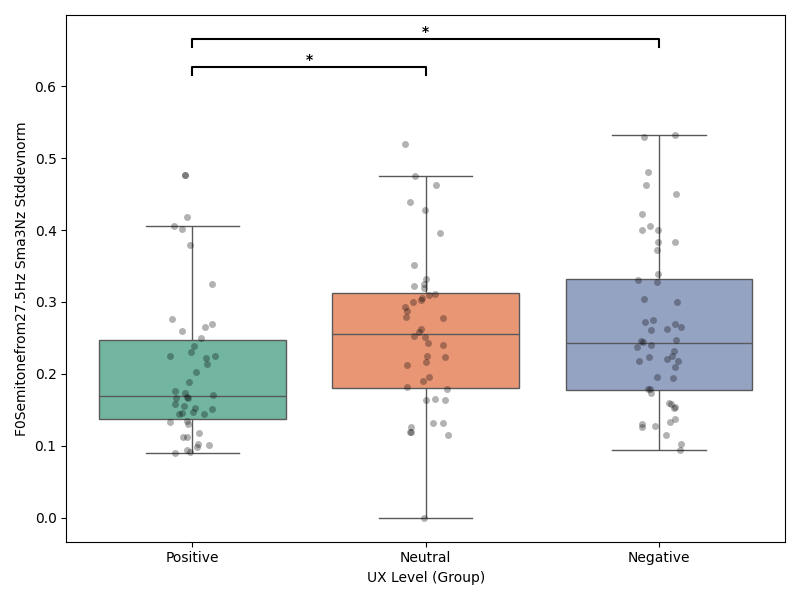}
        \caption{Pitch Variability (F0 semitones, std)}
        \label{fig:f0_std}
    \end{subfigure}
    \hfill
    \begin{subfigure}[b]{0.32\linewidth}
        \centering
        \includegraphics[width=\linewidth]{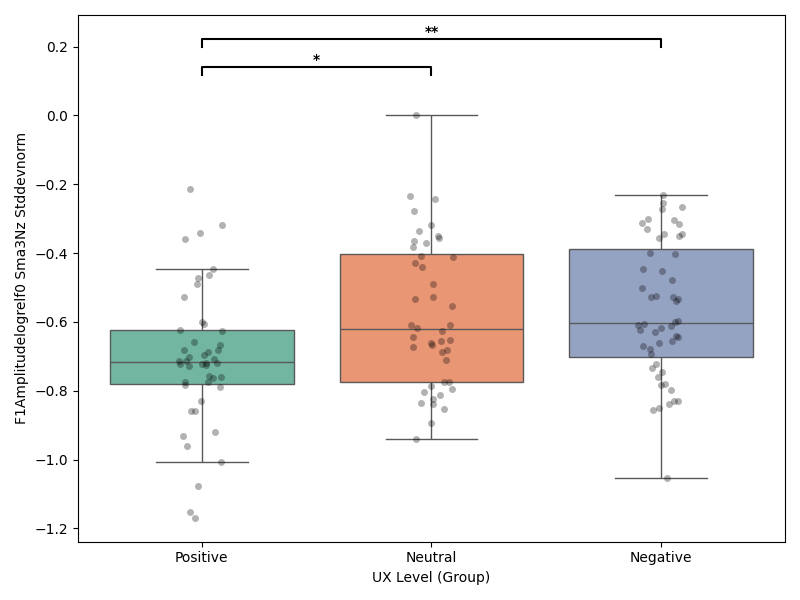}
        \caption{Formant Amplitude (F1–F0, std)}
        \label{fig:f1_std}
    \end{subfigure}

    \caption{\textcolor{black}{Distributions of six voice‐quality–related acoustic features across the three UX levels (Positive, Neutral, Negative). All features exhibited significant differences across groups based on Kruskal–Wallis tests with Dunn–Bonferroni post-hoc comparisons. Higher values of jitter, loudness variability, HNR variability, pitch variability, and F1 amplitude variability are consistently associated with lower UX quality, indicating increased vocal instability during more negative interactions.}}
    \Description{Six boxplots arranged in two rows showing distributions of intensity mean, jitter variability, loudness variability, harmonic-to-noise ratio variability, pitch variability in semitones, and formant amplitude variability across positive, neutral, and negative user experience levels, with significance annotations for group differences.}
    \label{fig:voice_quality_six}
\end{figure*}

Voice quality metrics, including jitter, shimmer, harmonic-to-noise ratio (HNR), and related measures, were extracted using the openSMILE toolkit. Our analysis showed that several of these parameters varied systematically with perceived UX quality. In particular, vocal intensity, fundamental frequency variability, loudness variability, jitter, HNR, and F1 amplitude all differed significantly across the three UX levels, indicating that users’ voice stability and timbral characteristics reliably tracked changes in interaction quality. As visualized in Figure~\ref{fig:voice_quality_six}, it summarizes the distributions of these features across the three UX levels (Positive, Neutral, Negative). Consistent with prior speech-science work linking perturbation to affective and cognitive strain, all six metrics showed significant differences across UX groups, as indicated by Kruskal–Wallis tests followed by Dunn–Bonferroni corrected comparisons.

Mean vocal intensity differed significantly across UX levels ($H = 10.71$, $p = .0047$, $\varepsilon^2 = .063$), with post-hoc tests revealing that Negative UX interactions produced reliably lower intensity than both Neutral ($p = .029$) and Positive UX ($p = .007$). This reduction in loudness is characteristic of guarded or effortful vocal behavior during problematic interactions. Fundamental frequency variability (F0 Semitone, SD) also differed significantly by UX level ($H = 10.41$, $p = .0055$, $\varepsilon^2 = .061$). Negative UX again showed significantly greater pitch variability compared to Positive UX ($p = .016$), suggesting more unstable pitch contours and increased speaking effort under frustration or uncertainty.
Loudness variability (Loudness, SD) exhibited significant group differences ($H = 9.36$, $p = .009$, $\varepsilon^2 = .053$). Negative UX elicited greater loudness fluctuation than Positive UX ($p = .012$), aligning with known markers of stress-related vocal instability. Jitter (cycle-to-cycle F0 perturbation) also showed a significant effect of UX level ($H = 9.79$, $p = .0075$, $\varepsilon^2 = .056$), with Negative UX displaying higher jitter than Neutral UX ($p = .006$). This reflects increased micro-instability in phonation when interactions are experienced as difficult or frustrating.
Harmonic-to-noise ratio (HNR) revealed significant differences across groups ($H = 9.67$, $p = .008$, $\varepsilon^2 = .055$). Recordings labeled as Negative UX showed significantly lower HNR than Positive UX ($p = .006$), indicating more turbulent, breathier vocal output — a pattern associated with strain or negative affect. Finally, F1 amplitude relative to F0 also differed significantly across UX categories ($H = 12.20$, $p = .0022$, $\varepsilon^2 = .073$). Negative UX interactions exhibited significantly higher (less negative) F1 amplitude than Positive ($p = .0025$) and Neutral UX ($p = .030$), suggesting alterations in articulatory stability under poorer interaction quality.

Taken together, these six voice-quality descriptors show a coherent pattern: as UX worsens, speech becomes more perturbed, less periodic, and more noise-dominated. Negative UX interactions consistently elicited greater pitch and loudness variability, increased jitter, reduced harmonic richness, and altered resonant characteristics. These results indicate that users’ frustration or cognitive strain is embodied in micro-perturbations of the voice, providing a robust set of physiological markers that voice interfaces could leverage for real-time UX sensing. The systematic differences illustrated in Figure~\ref{fig:voice_quality_six} reinforce the potential of voice quality as a fine-grained diagnostic signal for interaction breakdowns.

\subsection{Correlations with Subjective UX Dimensions (RQ2)}


Building on the UX-level analysis, we examined how Trust, Satisfaction, and Attractiveness relate to speech features at a continuous scale. Whereas the ANOVA tested categorical differences between UX groups, this section probes monotonic associations between subjective ratings and acoustic/NLP descriptors.



\begin{table}[htbp]
\centering
\caption{Correlations between perceived Trust and acoustic-prosodic features.}
\Description{Table showing Pearson correlation coefficients between perceived trust ratings and spectral acoustic features, including spectral bandwidth, spectral rolloff, zero-crossing rate, and spectral tilt slope, along with corresponding p-values and adjusted p-values.}
\label{tab:trust_correlations}

\resizebox{\linewidth}{!}{%
\begin{tabular}{lccc}
\toprule
\textbf{Spectral Feature} & \textbf{Pearson's \(r\)} & \textbf{\(p\)-value} & \textbf{Adjusted \(p\)} \\ 
\midrule
Spectral Bandwidth Mean & -0.251 & 0.003 & 0.357 \\
Spectral Rolloff Mean & -0.210 & 0.015 & 0.448 \\
Zero-Crossing Rate Mean & -0.201 & 0.019 & 0.448 \\
Spectral Tilt Slope & -0.196 & 0.023 & 0.448 \\
\bottomrule
\end{tabular}%
}

\end{table}

We extracted a comprehensive set of features with librosa, openSMILE, and Praat/Parselmouth, and derived paralinguistic proxies via an NLP pipeline. Across the three dimensions, we observed a consistent pattern of weak negative correlations for several spectral mean features, most notably spectral roll-off and spectral bandwidth. Although these effects did not remain significant after correction for multiple comparisons (see Tables~\ref{tab:trust_correlations}–\ref{tab:attractiveness_correlations}), their stable direction suggests a meaningful trend: as ratings decline, speakers’ voices show reduced high-frequency energy and narrower spectral dispersion. This acoustic profile, flatter and less resonant, may reflect lower engagement or positive affect, aligning with the perturbation patterns seen in the categorical analysis. While not conclusive on their own, these correlations reinforce the view that vocal output captures subtle facets of users’ subjective experience.
\begin{table}[t]
\centering
\caption{Correlations between perceived Satisfaction and acoustic spectral features.}
\Description{Table reporting Pearson correlation coefficients between perceived satisfaction ratings and selected spectral acoustic features, including spectral rolloff, spectral bandwidth, and spectral centroid means, along with corresponding p-values and adjusted p-values.}
\label{tab:satisfaction_correlations}

\resizebox{\linewidth}{!}{%
\begin{tabular}{lccc}
\toprule
\textbf{Spectral Feature} & \textbf{Pearson's \(r\)} & \textbf{\(p\)-value} & \textbf{Adjusted \(p\)} \\ 
\midrule
Spectral Rolloff Mean & -0.254 & 0.003 & 0.262 \\
Spectral Bandwidth Mean & -0.234 & 0.007 & 0.318 \\
Spectral Centroid Mean & -0.219 & 0.011 & 0.318 \\
\bottomrule
\end{tabular}%
}

\end{table}

\begin{table}[ht]
\centering
\caption{Correlations between perceived Attractiveness and acoustic spectral features. }
\Description{Table reporting Pearson correlation coefficients between perceived attractiveness ratings and selected spectral acoustic features, including spectral rolloff, spectral bandwidth, and spectral centroid means, along with corresponding p-values and adjusted p-values.}
\label{tab:attractiveness_correlations}

\resizebox{\linewidth}{!}{%
\begin{tabular}{lccc}
\hline
\textbf{Spectral Feature} & \textbf{Pearson's \(r\)} & \textbf{\(p\)-value} & \textbf{Adjusted \(p\)} \\ 
\hline
Spectral Rolloff Mean & -0.258 & 0.002 & 0.218 \\
Spectral Bandwidth Mean & -0.231 & 0.007 & 0.284 \\
Spectral Centroid Mean & -0.231 & 0.007 & 0.284 \\
\hline
\end{tabular}
}
\end{table}

Beyond attractiveness, trust, and satisfaction, we extended our analysis to key conversational UX dimensions in VAs: comprehensibility, response quality, and response behavior~\cite{klein2024exploring, klein2024flexible}. A complete synthesis of these speech–UX relationships, including effect sizes, statistical significance, and individual differences, is available in Appendix~\ref{speech_ux}. This analysis reveals, for example, that response quality degradation correlates strongly with increased interaction duration ($\varepsilon^2$ = .090) and elevated harmonic-to-noise variability ($\varepsilon^2$ = .060).

\subsection{Predicting UX Levels from Speech (RQ3)}

To assess the practical value of speech as a passive UX signal, we framed UX recognition as a three-way classification task, predicting \textit{Positive}, \textit{Neutral}, and \textit{Negative} UX labels directly from acoustic features. Building on the descriptive analyses in Sections~\ref{ux_classification}, we constructed feature sets from low-level descriptors capturing prosody, spectral shape, social speech behavior, and voice quality (e.g., intensity, F0, jitter, shimmer, spectral centroid, speech activity, and pause structure). All models were evaluated using stratified 5-fold cross-validation to preserve the empirical class distribution while providing robust estimates of generalization performance.

We first compared a set of classical machine-learning baselines using our extracted acoustic descriptors as input. After z-normalization, we trained $k$-Nearest Neighbors (KNN), a Support Vector Machine (SVM with RBF kernel), Random Forests, and XGBoost, performing inner cross-validated grid search for hyperparameter tuning. Across folds, the SVM achieved the best performance with a mean accuracy of $76.47\%$ ($SD = 1.55$), clearly outperforming KNN ($69.50\%$, $SD = 2.60$), Random Forests ($68.08\%$, $SD = 2.74$), and XGBoost ($67.25\%$, $SD = 2.46$). On the final evaluation fold, the SVM reached $74.56\%$ accuracy, with class-wise F1-scores of $0.71$ (Positive), $0.73$ (Neutral), and $0.80$ (Negative), indicating that purely acoustic features are already sufficient to distinguish UX levels well above the chance level of $33.3\%$.

\begin{figure}[t]
    \centering
    \includegraphics[width=\linewidth]{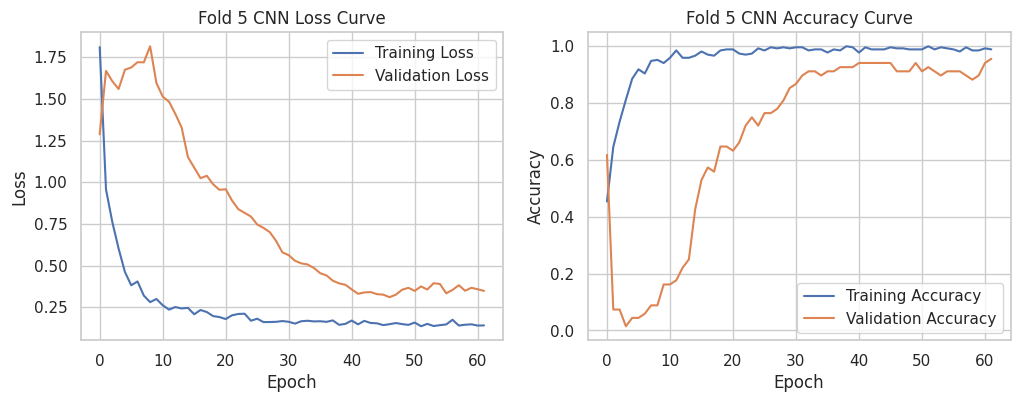}
    \caption{Training and validation learning curves for the 1D--CNN UX classifier (Fold 5). The model demonstrates stable convergence with early stopping.}
    \Description{Line plot showing training and validation accuracy curves across epochs for the 1D convolutional neural network UX classifier, illustrating model convergence and early stopping behavior.}
    \label{fig:cnn_learning_curves}
\end{figure}

To model higher–order patterns in the acoustic feature space, we trained a 1D convolutional neural network (1D–CNN) using both our extracted acoustic descriptors and the eGeMAPS functional set. Each input feature vector was normalized via in-model Batch Normalization and processed through a lightweight convolutional block comprising 64 filters (kernel size = 5), ReLU activation, max pooling, and $L_2$ regularization. The convolutional representation then fed into a dropout-regularized dense layer and a final softmax classification layer. To mitigate data sparsity and improve robustness, we applied feature-space augmentation within each training fold by injecting Gaussian noise and perturbing pitch-related dimensions (F0, pitch, jitter). This procedure effectively doubled the training data. 
Using 5-fold stratified cross-validation with early stopping, the 1D-CNN achieved a mean accuracy of $0.76$ (SD = 0.03), with fold-level performance ranging from $0.72$ to $0.80$. This accuracy closely matches that of the strongest classical baseline (SVM: $0.76$ mean accuracy), indicating that UX-related acoustic cues are learnable by both shallow and moderately deep architectures. Learning curves (Figure \ref{fig:cnn_learning_curves}) show stable convergence without signs of overfitting, suggesting that the model captures consistent, non-linear structure in the mapping from speech to UX states.

These predictive results offer a concrete proof-of-concept for speech-based, real-time UX monitoring in voice interfaces. Both the conventional model (SVM) and the deep CNN architecture reliably discriminate Positive, Neutral, and Negative UX from a single voice interaction, indicating that UX-related information in speech is not only statistically significant but also predictive at the level required for practical applications.

\section{Discussion}
\label{discussion}

We conducted a comprehensive evaluation of UX with VAs by examining interactions across functional, collaborative, and playful scenarios. By integrating self-reported measures (mood, stress, UEQ+) with speech-acoustic and linguistic features extracted from user utterances, we adopted a mixed-methods approach. This methodology allowed us to identify not only categorical differences between assistants but also continuous relationships between subjective experience and vocal behavior, providing a more holistic understanding of UX in voice-based interactions.

\subsection{Reflections on Our Approach and Contribution to Prior Work}

Our findings provide strong evidence that speech features recorded during VUI interactions can systematically differentiate UX quality in interpretable ways. Across spectro-temporal, social–behavioral, and voice-quality measures, we observed coherent patterns consistent with speech science literature: smoother, more stable speech during positive interactions, and brighter, more perturbed, or unstable speech during negative experiences. While these patterns align with prior work linking prosody to stress, frustration, and cognitive load \cite{kirkland2022s,munoz2020fundamental,ang2002prosody,song2021frustration}, our contribution lies in demonstrating them within \emph{in-situ, task-oriented} interactions with functional conversational agents — moving beyond retrospective surveys, acted corpora, or laboratory emotion-elicitation setups.

Moreover, by designing three VA personas that jointly manipulated output length, error-handling strategy, response latency, personality, and voice quality, we intentionally elicited a wide range of UX. This contrasts with prior VUI research, which often varies only a single design factor within scripted or tightly controlled dialogues \cite{porcheron2018voice,dutsinma2022systematic}. Our approach reveals how multiple design choices interact to shape users’ vocal behavior during authentic, real-time conversation, providing a more valid basis for adaptive VUI design.

Our results contribute three key advances to VA UX research. First, we demonstrate that multiple UX constructs, including Trust, Attractiveness, Satisfaction, Comprehensibility, and Response Behavior, are reflected not only in post-hoc surveys but also in users’ spontaneous speech during interaction. This extends prior work that primarily correlates UX with post-task ratings \cite{frison2019ux,klein2024exploring,kocaballi2019understanding}, by showing that speech carries a rich, multidimensional behavioral signal that mirrors users’ evolving experience with the system in real time. Second, by integrating acoustic, prosodic, and social speech features, we identify a coherent acoustic signature of negative UX: brighter and noisier spectra, greater instability in vocal control, and more effortful conversational behavior. This pattern aligns with theories of speech under stress and cognitive strain \cite{van2018voice,dahl2021changes,ang2002prosody,caralt2025stupid}, but we validate it within the context of everyday VA use — a domain where such models have rarely been tested at scale in naturalistic settings. Third, we provide evidence that UX-related changes in speech emerge naturally, without users being instructed to modify their speaking style. Unlike studies that rely on elicited emotions \cite{swoboda2022comparing,ma2025advancing} or pre-scripted dialogues \cite{allison2018design,branham2019reading}, our approach captures UX dynamics during task-driven interactions, demonstrating the feasibility of \emph{passive, real-time UX sensing}. This positions our work alongside efforts in adaptive conversational interfaces \cite{dhakal2019near,porcheron2018voice} but with stronger empirical grounding in ecological speech data and clearer interpretability.

These findings carry direct implications for how VAs could detect and respond to UX issues in real time. For example, instability in spectral flux or loudness, which we found to be strong markers of UX degradation, could be continuously monitored, enabling VAs to recognize emerging frustration before users explicitly express it. Similarly, reductions in engagement rate or speech activity may signal disengagement or task struggle. Because these cues are non-lexical, they offer a \emph{privacy-preserving pathway} to on-device UX modeling without analyzing speech content. Furthermore, unlike prior work that assesses UX through post-study interviews \cite{ma2025measuring,fan2021older}, our system enables \emph{real-time UX detection}, opening possibilities for immediate system adaptation and more dynamic evaluation paradigms. This shift from retrospective to concurrent measurement aligns with growing interest in continuous, unobtrusive UX assessment in HCI.

\subsection{Clear Differentiation between VAs}

Across measures, VA~A consistently outperformed VA~B and VA~C. Participants reported better mood and lower stress after interacting with VA~A, and rated it highest on efficiency, usefulness, trust, and satisfaction. VA~C received the lowest scores and frequent critical comments (e.g., recognition failures, slow or garbled output, irrelevant responses). VA~B occupied the middle ground—acceptable or even enjoyable in some scenarios, but penalized for noise, repetition, and comprehension issues. The rank ordering (A $>$ B $>$ C) in the post‐study comparison mirrors these patterns, indicating that users are sensitive to reliability, fluency, and contextual relevance.

These findings echo earlier work showing that even small variations in latency, repair strategies, or persona design can produce substantial differences in perceived system quality~\cite{funk2020usable,niculescu2013making}.  Our results further suggest that these design-driven differences are reflected directly in users’ vocal behavior. Meanwhile, individual users may still experience the same assistant differently, indicating that UX is shaped not only by system performance but also by personal expectations, speaking styles, and interaction preferences.

\subsection{Speech Mirrors UX — Acoustic Evidence}
The acoustic analysis offers objective, behavioral corroboration of the subjective reports. Our statistical analyses revealed significant group differences for several variability measures, including the standard deviation of Root-Mean-Square Energy ($\mathrm{RMSE_{std}}$), Spectral Centroid ($\mathrm{SC_{std}}$), Zero-Crossing Rate ($\mathrm{ZCR_{std}}$), and Spectral Bandwidth ($\mathrm{BW_{std}}$). Poorer UX consistently produced flatter vocal dynamics: reduced variability in energy and spectral structure, alongside diminished high-frequency emphasis.

This acoustic profile aligns with known signatures of reduced engagement, heightened cognitive load, and negative affect in speech science~\cite{kirkland2022s,ang2002prosody}. In combination with self-reported decreases in mood and trust, these findings suggest that speech acoustics provide a reliable, non-invasive window into user state during naturalistic interaction.




\subsection{Continuous Links to Subjective Dimensions}
Beyond categorical differences, correlational analyses revealed continuous relationships between speech features and subjective UX dimensions like \emph{Trust}, \emph{Satisfaction}, and \emph{Attractiveness}. We observed a pattern of small, predominantly negative associations with spectral means (e.g., spectral roll-off, bandwidth, centroid): as user ratings decreased, high-frequency energy and spectral dispersion tended to drop. Although these correlational effects were modest and did not survive strict correction for multiple comparisons, the consistent direction of effects converges with the results from our group-level analyses. This convergence suggests a coherent trend: vocal output captures subtle, continuous aspects of how users feel about a system, even when individual features explain only a small portion of the variance independently.

\subsection{Assistant Effects Dominate Scenario Effects}
While the interaction scenario (trip planning, storytelling, fortune telling) had a modest effect on mood ratings, the assistant’s persona was the dominant factor determining UX. This finding has an important practical implication: speech-derived vocal cues appear to be more diagnostic of assistant quality and the resulting user state than of the task genre itself. This strengthens the case for using vocal analytics as a general-purpose tool for VA evaluation and real-time adaptation, even across diverse interaction contexts.



\subsection{Implications for UX Evaluation and Measurement}

The broader methodological implication of this work is that speech offers a feasible and informative modality for continuous UX assessment. Unlike retrospective questionnaires, which capture only a summary of the user’s experience after an interaction, speech is produced constantly and reflects moment-by-moment fluctuations in engagement, trust, and satisfaction. This positions speech as a promising complement to existing UX methods, particularly in contexts like VUIs where interaction is already mediated through the user’s voice. The fact that many of the most reliable indicators in our study were non-lexical means that such sensing could be achieved in ways that preserve user privacy by avoiding any analysis of linguistic content.

Another implication concerns the temporal granularity of UX understanding. Acoustic features allow systems and researchers to detect turning points in the interaction, for example, the point at which latency becomes frustrating or at which an unhelpful response creates uncertainty. This capability moves the field toward a more dynamic and fine-grained notion of UX, opening possibilities for evaluating systems in situ without interrupting the user’s task flow. By demonstrating that UX-related vocal patterns are consistent across multiple personas and interaction scenarios, our findings offer evidence that these signals are not idiosyncratic but instead reflect generalizable characteristics of human–AI interaction.

\subsection{Implications for the Design of Adaptive Voice Assistants}

The results also carry important implications for the design of future adaptive VAs. The consistency with which user speech responded to system latency, persona behaviour, and error-handling strategies suggests that VAs could monitor subtle changes in vocal behaviour to infer when an interaction is deteriorating. Such monitoring would allow systems to adjust their behaviour proactively, for example, by simplifying responses, slowing delivery, or offering clarification when signs of tension or confusion appear in the user’s voice. This reinforces the growing recognition in HCI that adaptive, user-state–aware systems can mitigate conversational breakdowns and enhance trust in AI-driven interactions.

Moreover, the absence of systematic differences between demographic groups in our dataset suggests that the acoustic markers identified here may have broad applicability. At the same time, individual differences in vocal style mean that adaptive systems should track relative changes over time rather than relying on fixed thresholds. Our findings thus highlight the need for adaptive baselines and personalized models, while also demonstrating that universal patterns exist at the level of group behaviour.

\section{Limitations and Future Work}
\subsection{Limitations}
This study demonstrates meaningful relationships between vocal characteristics and UX, yet several limitations should be considered. First, the ecological variability of remote audio data collection introduced heterogeneity in recording conditions. Participants used personal devices with varying microphone quality, ambient noise environments, and platform-specific audio processing algorithms. While enhancing real-world validity, this variability may have introduced noise into acoustic measurements such as spectral energy and frequency characteristics.

Another limitation concerns our ground truth UX labels combined with UEQ+ benchmark framework by Meiners et al.~\cite{meiners2024benchmark}, potentially conflating distinct experiential constructs such as satisfaction, trust, and usability. This aggregation, while practical, may have obscured more nuanced relationships between specific UX dimensions and particular vocal features. Additionally, the concurrent variation of multiple assistant characteristics (e.g., recognition accuracy, latency, voice personality) limits our ability to make causal claims about which specific design factors drive which vocal changes.

Moreover, our participants were primarily drawn from English-speaking populations. Vocal characteristics, speech patterns, and user expectations may vary across linguistic, cultural, and demographic groups, potentially limiting the generalizability of our findings. Future work should explicitly test cross-cultural validity and incorporate fairness considerations into model development.

Finally, the inherent entanglement of prosodic features with linguistic content and conversational context presents an ongoing challenge. While we analyzed acoustic features comprehensively, disentangling the specific contributions of UX from other factors affecting speech production requires additional methodological innovations. Furthermore, this complexity may also influence the performance of our machine learning methods, as our feature selection was necessarily limited to the speech and paralinguistic features identified as relevant in our analysis. In our classification approach, we employed KNN, SVM, Random Forests, XGBoos and DNN models combined with these curated features, achieving strong but potentially improvable performance. This methodology may represent a limitation, as state-of-the-art deep learning architectures — such as transformer-based models, self-supervised speech representations (e.g., Wav2Vec 2.0~\cite{baevski2020wav2vec}, HuBERT~\cite{hsu2021hubert}), or end-to-end learning frameworks — could capture more nuanced, hierarchical patterns directly from raw or minimally processed audio signals. Future work should explore these advanced methods, which may improve classification accuracy and generalization while reducing reliance on manual feature engineering. Additionally, multimodal approaches that jointly model acoustic, linguistic, and dialogue context features could further enhance the robustness and interpretability of UX prediction models.

\subsection{Future Work}

Building on our findings, we identify several important directions for future research. First, our study varied multiple design factors simultaneously through the three VA personas. Although this approach allowed us to observe how persona-wide differences manifest in users’ speech, it limits our ability to isolate the causal contribution of specific design elements. Future controlled experiments should manipulate factors such as error rate, response latency, and vocal prosody independently to establish clearer causal relationships between system behaviors and the vocal biomarkers of UX. Such work would offer more precise design guidance for voice interfaces.
Second, the present study involved a relatively homogeneous participant group and a single linguistic context.
Because voice production varies across cultures, languages, and demographic groups, future research should examine the generalizability of these findings in more diverse populations. Cross-cultural validation studies, domain-adaptive acoustic modeling, and explicit evaluation across demographic subgroups will all be essential for ensuring ecological validity and fairness in speech-based UX assessment.
Third, our analyses focused on acoustic and behavioral speech features, but UX is inherently multidimensional. Future work should explore multimodal frameworks that integrate acoustic signals with linguistic content (e.g., disfluencies, sentiment, repair utterances) and interactional structure (e.g., turn-taking patterns, self-repairs, hesitation sequences). Such integrated models may better disentangle the relationship between UX, speech production, and conversational dynamics, advancing both interpretability and predictive accuracy.
Finally, translating speech-based UX sensing into real-world applications raises several practical challenges. Implementing continuous, on-device assessment will require efficient streaming feature extraction, reliable smoothing or aggregation mechanisms that prevent over-reacting to transient vocal fluctuations, and strong privacy protections. Exploring architectures that preserve user autonomy, such as federated processing or minimization of linguistic content, will be critical for enabling ethically responsible deployments.
Addressing these limitations would deepen theoretical understanding of how UX is expressed through speech while supporting the development of more adaptive, user-centered voice interaction systems capable of responding to user state in real time.

\section{Conclusion}
This paper provides compelling evidence that speech offers a non-invasive and reliable lens through which to understand UX with VAs. Across personas and scenarios, lower UX was consistently reflected in users’ vocal behaviour: increased within-utterance variability in energy and spectral structure, reduced activity and engagement, and degraded voice quality. These patterns form complementary markers of cognitive strain, conversational misalignment, and emerging user frustration. Although correlations with individual UX dimensions were modest, they were systematic and coherent across participants, suggesting that speech captures generalizable signatures of UX rather than idiosyncratic variation.
Our modeling results further demonstrate that a compact and interpretable feature set — spanning prosody and timing, voice quality, time–frequency structure, and transcript-derived indicators — can classify UX levels with accuracies of up to 80\%. Together, these findings highlight the potential of speech-based sensing to move beyond the limitations of post-hoc surveys and toward passive, turn-level, in-situ UX estimation. This shift opens a pathway for VUIs that not only measure experience but also adapt to it dynamically.
In future work, we aim to strengthen the robustness and generalizability of our machine learning models across contexts, tasks, and populations, link detected vocal cues to concrete and designable system responses, and develop privacy-preserving, on-device processing pipelines.
Overall, these advances will support the development of VAs that are not only more capable, but also more sensitive and responsive to how people actually feel while interacting with them.

\begin{acks}
This work was funded by the Deutsche Forschungsgemeinschaft (DFG, German Research Foundation) -- SPP 2199 -- in the Project TransforM (425412993), the Research Council of Norway under project number 326907, the National Natural Science Foundation of China 
(Grant No.W2533169) and the SUSTech Presidential Postdoctoral Fellowship.
Also, we gratefully acknowledge funding from UiB (University of Bergen), which supported the research activities reported in this paper.
\end{acks}

\bibliographystyle{ACM-Reference-Format}
\bibliography{sample-base}

\clearpage
\appendix

\section{Post-Study Structured Interview Methodology and Analysis}
\label{structured_interview}
To capture open-ended, spoken reflections on the three voice assistant personas, participants completed a brief structured interview at the conclusion of the experimental session. This section details the interview procedure, analytical method, and the key qualitative themes derived from participant responses.
\subsection{Interview Procedure and Design}
The interview was delivered via a dedicated “interview VA” interface (Figure~\ref{fig:interview_va}).
This VA presented four fixed, scripted questions in the same order to all participants, pausing after each to record the participant’s spoken response. The questions were:


\begin{enumerate}
    \item How are you feeling now, after finishing our study?
    \item How was your experience with each voice assistant?
    \item Which scenario did you prefer the most?
    \item Is there anything else you would like to share about your experience?
\end{enumerate}

The interview was strictly structured; the VA did not ask adaptive follow-up questions. This design ensured consistency across all 49 participants while collecting spontaneous, voice-based feedback that maintained the study’s interactive modality.

\subsection{Data Processing and Analytical Method}
All interview audio recordings were automatically transcribed using OpenAI’s Whisper model and manually verified for accuracy. The resulting textual data was analyzed using inductive thematic analysis following the six-phase framework by Braun and Clarke~\cite{braun2006using}. The process involved (1) familiarization with the data, (2) generating initial codes, (3) searching for themes, (4) reviewing and refining themes, (5) defining and naming themes, and (6) producing the final report. This analysis identified four central themes characterizing participants’ perceptions of voice assistant interaction quality.

\subsection{Key Qualitative Themes}
The thematic analysis revealed consistent patterns in how participants evaluated their experiences, which are summarized below with illustrative participant quotations.

\paragraph{Frustration Driven by System Unreliability.}
Participants consistently expressed strong negative reactions when assistants were unresponsive, misunderstood commands, or exhibited audio glitches. These technical failures often led to task abandonment. For instance, P19 stated, “The last one didn’t even work… it kept on saying, ‘I didn’t get that.’ Couldn’t even use it.” Similarly, P27 reported, “Two of them were incredibly frustrating… I could barely get any sort of response from them.” These accounts align with the significantly lower UEQ+ scores in Dependability and Perspicuity for Personas B and C.

\paragraph{Positive Engagement Linked to Conversational Fluency.} 
Interactions marked by accurate understanding, prompt responses, and coherent dialogue were described as enjoyable and satisfying. The storytelling persona (Persona A) was frequently praised for its collaborative and creative nature. P10 remarked, “The first voice assistant… was amazing. I loved the creativity and the lack of errors,” and P49 noted, “The second one was super fun. I enjoyed it… it was really cool.” These reports correlate with Persona A’s higher UEQ+ ratings in Stimulation, Novelty, and Satisfaction.

\paragraph{Annoyance from Artificial and Non‑Adaptive Behavior.}
Many participants critiqued the synthetic vocal quality and generic, scripted nature of responses, which reduced perceived competence and trust. Descriptors like “robotic” and “artificial” were common. P12 commented, “It felt like way too artificial and glitchy… like I was talking to a broken computer,” while P13 observed, “The speech is very robotic. The answers were predictable and sometimes slow.” This theme underscores the importance of vocal expressiveness, reflected in lower Attractiveness scores for Personas B and C.

\paragraph{Scenario‑Specific Expectations Shape Preference.}
Participant preferences were closely tied to whether an assistant’s capabilities matched the demands of the task. Those valuing utility praised the functional trip planner when it worked reliably. P20 said, “The trip planning one actually listened to my requests,” and P45 stated, “The trip planner was great… it did everything it was supposed to do.” In contrast, participants seeking engagement favored the creative storytelling assistant. P2 described it as “seamless in its interaction,” and P44 called it “actually interactive.” The fortune teller elicited mixed reactions, valued more for novelty than utility, with P37 noting it was “mildly interesting” but gave “generic responses.”

\subsection{Synthesis and Connection to Quantitative Findings}
The qualitative insights from these interviews provide explanatory depth to the quantitative results presented in Section~\ref{questionnaire_label} and~\ref{speech_analysis}. They confirm that low UEQ+ scores and perturbed speech features often corresponded with participant reports of frustration and system failure. Conversely, high ratings and stable acoustic profiles aligned with descriptions of fluency and enjoyment. This triangulation of methods strengthens the validity of our findings, illustrating that voice assistant UX is co-constructed by technical reliability, conversational fluidity, vocal character, and scenario alignment. These themes directly inform the design implications discussed in Section~\ref{discussion}.

\section{Additional Persona Interaction Examples}
\label{appendix:persona-examples}

\begin{table*}[htbp]
\renewcommand{\arraystretch}{1.35}
\caption{Example utterances illustrating response style and error handling for each VA persona, based on the design levers in Table~\ref{table1}.}
\Description{Table presenting example voice assistant responses for three personas across a weather query task and an error-handling scenario, illustrating differences in response style, emotional tone, and clarification behavior.}
\label{table:persona-examples-table}
\centering
\begin{small}
\begin{tabular}{|p{2.1cm}|p{5.1cm}|p{5.2cm}|}
\hline
\bfseries Persona & \bfseries Task Response Example (Weather Query) & \bfseries Error-Handling Example \\
\hline

\textbf{VA-A (Warm, Competent)} &
``Of course! Checking… Great news --- Rome will be sunny and warm, around 30\(^\circ\)C with no rain. Perfect sightseeing weather!'' &
``Just to confirm, did you mean `remind me to email the itinerary on Thursday at 9 AM'?'' \\

\hline

\textbf{VA-B (Flat, Emotionless)} &
``Forecast for Rome: sunny, around 30\(^\circ\)C. No rain.'' &
``Reminder created.'' (No clarification, even if misrecognized) \\

\hline

\textbf{VA-C (Irritated, Inconsistent)} &
(After long delay + degraded audio) ``Ugh… hold on… It’ll be hot. Thirty degrees. Sunny. There — you have your weather.'' &
(After long pause) ``Setting your reminder for Thursday at \textbf{9 PM}…'' (User corrects) ``Fine… changed to \textbf{9 AM}. Reminder set.'' \\

\hline
\end{tabular}
\end{small}
\end{table*}

To illustrate how the design levers described in Table~\ref{table:persona-examples-table} manifest in actual interactions, we provide representative utterances for each VA persona. These examples highlight differences in tone, verbosity, latency, and error-handling behavior, including VA C’s incorrect first response followed by a reluctant correction.








\section{Participant Demographics}
\label{tab:participants}


We recruited 49 participants (21 women, 28 men) via Prolific. Participants ranged in age from 20 to 50 years (\(M = 35.2\), \(SD = 8.1\)). Most participants reported prior experience with voice assistants (\(n = 44\), 89.8\%), while five participants (10.2\%) had never used a VA. Among experienced users, self-reported usage frequency varied: eight participants (18.2\%) used VAs very often, 19 (43.2\%) often, 13 (29.5\%) sometimes, and four (9.1\%) rarely.
Participants also self-rated their general talkativeness on a 7-point Likert scale (\(M = 3.3\), \(SD = 1.1\)). A detailed breakdown of individual demographics, VA usage background, and talkativeness ratings is provided in Table~\ref{tab:participants-split}. 

\begin{table*}[t]
\centering
\small
\caption{Participant demographics and voice assistant (VA) usage background (N=49). Usage frequency categories: VO = Very Often, O = Often, S = Sometimes, R = Rarely. Talkativeness rated from 1 (Very Quiet) to 7 (Very Talkative).}
\Description{Two side-by-side tables listing participant demographics including age, gender, prior voice assistant use, usage frequency, and self-rated talkativeness for all 49 participants.}
\label{tab:participants-split}

\renewcommand{\arraystretch}{1.1}
\setlength{\tabcolsep}{6pt}

\begin{tabular}{@{} c c l l c c @{}}
\toprule
\textbf{PID} & \textbf{Age} & \textbf{Gender} & \textbf{VA Use} & \textbf{Usage Freq.} & \textbf{Talk.} \\
\midrule
P1 & 34 & F & Yes & R & 3 \\
P2 & 50 & F & Yes & O & 3 \\
P3 & 37 & M & Yes & O & 4 \\
P4 & 24 & M & Yes & R & 2 \\
P5 & 26 & M & Yes & R & 3 \\
P6 & 31 & M & Yes & VO & 4 \\
P7 & 30 & M & Yes & VO & 5 \\
P8 & 33 & M & Yes & S & 4 \\
P9 & 38 & M & Yes & O & 3 \\
P10 & 20 & M & Yes & O & 4 \\
P11 & 37 & M & Yes & O & 4 \\
P12 & 33 & F & Yes & S & 5 \\
P13 & 50 & F & Yes & O & 3 \\
P14 & 26 & F & No & -- & 1 \\
P15 & 39 & F & Yes & VO & 4 \\
P16 & 37 & M & Yes & O & 3 \\
P17 & 40 & F & Yes & R & 4 \\
P18 & 29 & F & Yes & S & 2 \\
P19 & 26 & M & Yes & S & 3 \\
P20 & 30 & M & Yes & S & 3 \\
P21 & 38 & M & Yes & O & 2 \\
P22 & 34 & F & Yes & O & 4 \\
P23 & 25 & F & Yes & R & 2 \\
P24 & 31 & F & No & -- & 4 \\
P25 & 34 & M & Yes & O & 3 \\
\bottomrule
\end{tabular}
\hspace{0.8cm}
\begin{tabular}{@{} c c l l c c @{}}
\toprule
\textbf{PID} & \textbf{Age} & \textbf{Gender} & \textbf{VA Use} & \textbf{Usage Freq.} & \textbf{Talk.} \\
\midrule
P26 & 25 & F & Yes & S & 2 \\
P27 & 27 & M & Yes & R & 2 \\
P28 & 37 & F & Yes & VO & 5 \\
P29 & 44 & F & No & -- & 2 \\
P30 & 46 & M & Yes & O & 4 \\
P31 & 41 & F & Yes & O & 5 \\
P32 & 24 & M & Yes & VO & 5 \\
P33 & 37 & F & Yes & O & 5 \\
P34 & 50 & M & Yes & O & 3 \\
P35 & 27 & M & No & -- & 3 \\
P36 & 42 & M & Yes & S & 3 \\
P37 & 46 & F & Yes & S & 2 \\
P38 & 33 & M & Yes & O & 2 \\
P39 & 35 & M & Yes & S & 4 \\
P40 & 42 & M & Yes & S & 2 \\
P41 & 46 & F & Yes & S & 3 \\
P42 & 34 & F & Yes & S & 4 \\
P43 & 43 & M & Yes & O & 3 \\
P44 & 23 & M & Yes & R & 3 \\
P45 & 32 & M & Yes & S & 3 \\
P46 & 32 & F & Yes & O & 5 \\
P47 & 50 & M & Yes & O & 3 \\
P48 & 27 & M & Yes & VO & 2 \\
P49 & 46 & M & Yes & O & 2 \\
\bottomrule
\end{tabular}

\end{table*}

\section{Detailed Statistical Results for Paralinguistic and Lexical Speech Features}
\label{nlp_features_analysis}

\begin{figure*}[t]
    \centering

    \begin{subfigure}[b]{0.32\linewidth}
        \centering
        \includegraphics[width=\linewidth]{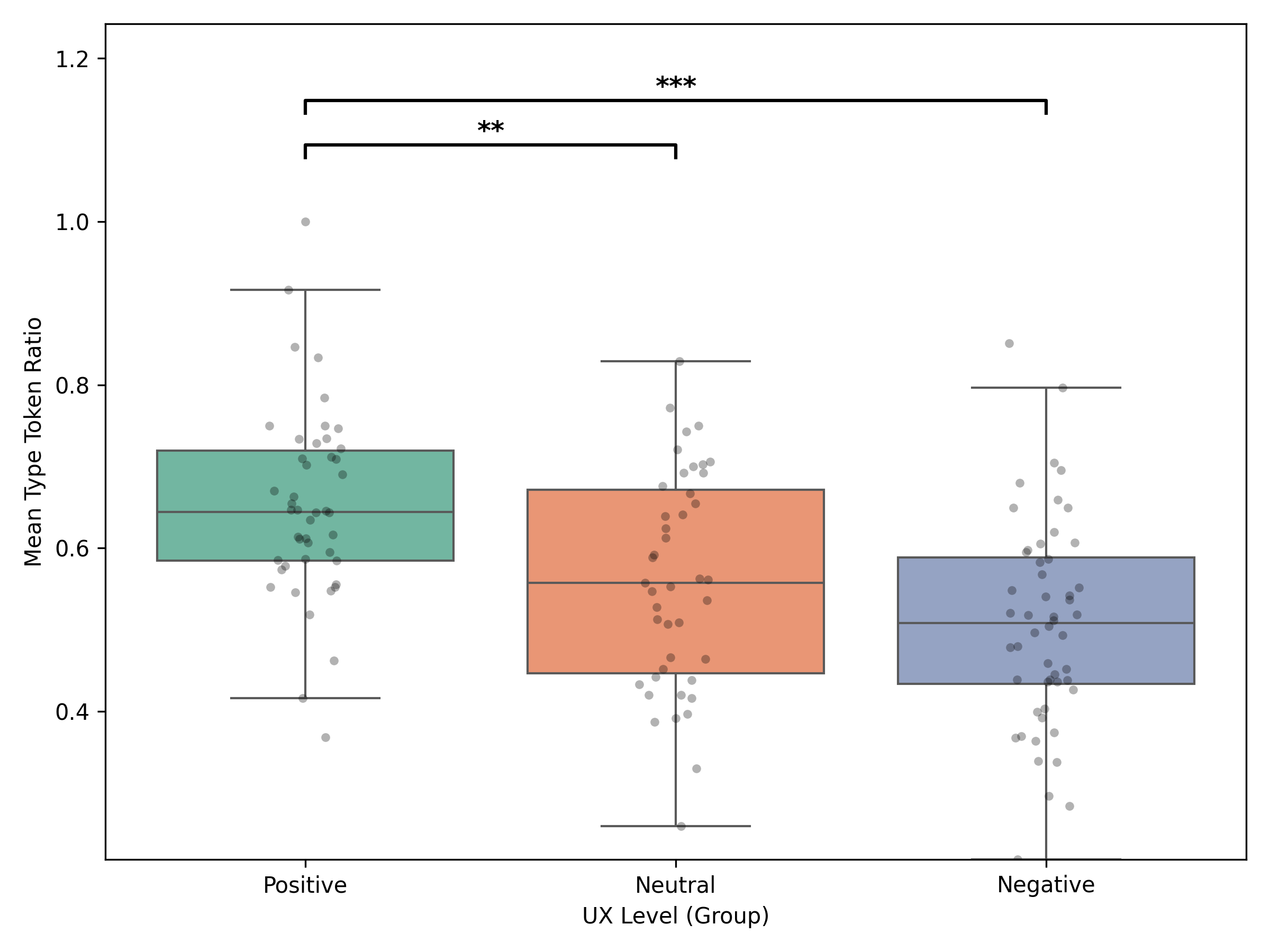}
        \caption{Mean type--token ratio}
        \label{fig:mtt_ratio}
    \end{subfigure}
    \hfill
    \begin{subfigure}[b]{0.32\linewidth}
        \centering
        \includegraphics[width=\linewidth]{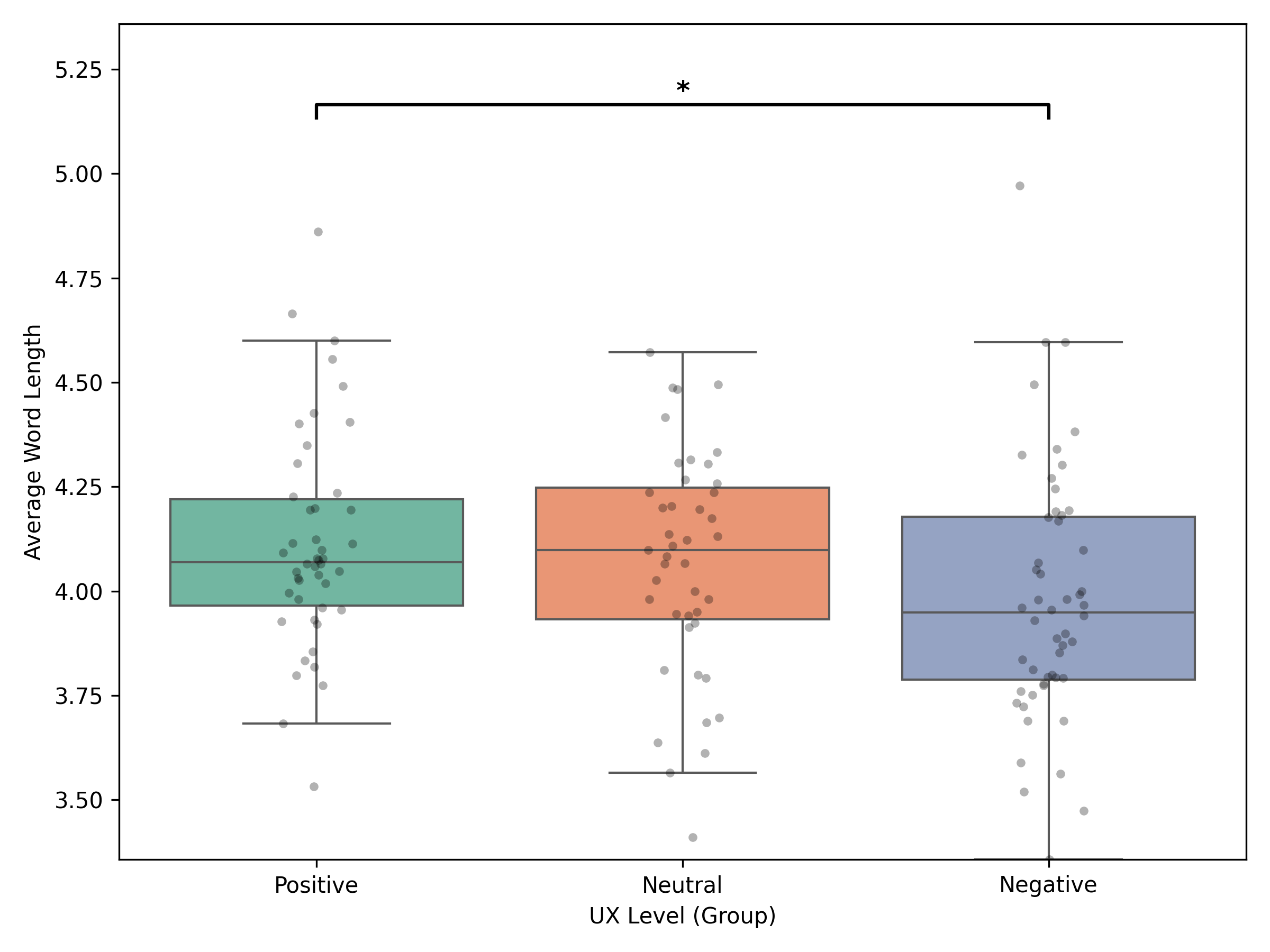}
        \caption{Average word length}
        \label{fig:avg_word_length}
    \end{subfigure}
    \hfill
    \begin{subfigure}[b]{0.32\linewidth}
        \centering
        \includegraphics[width=\linewidth]{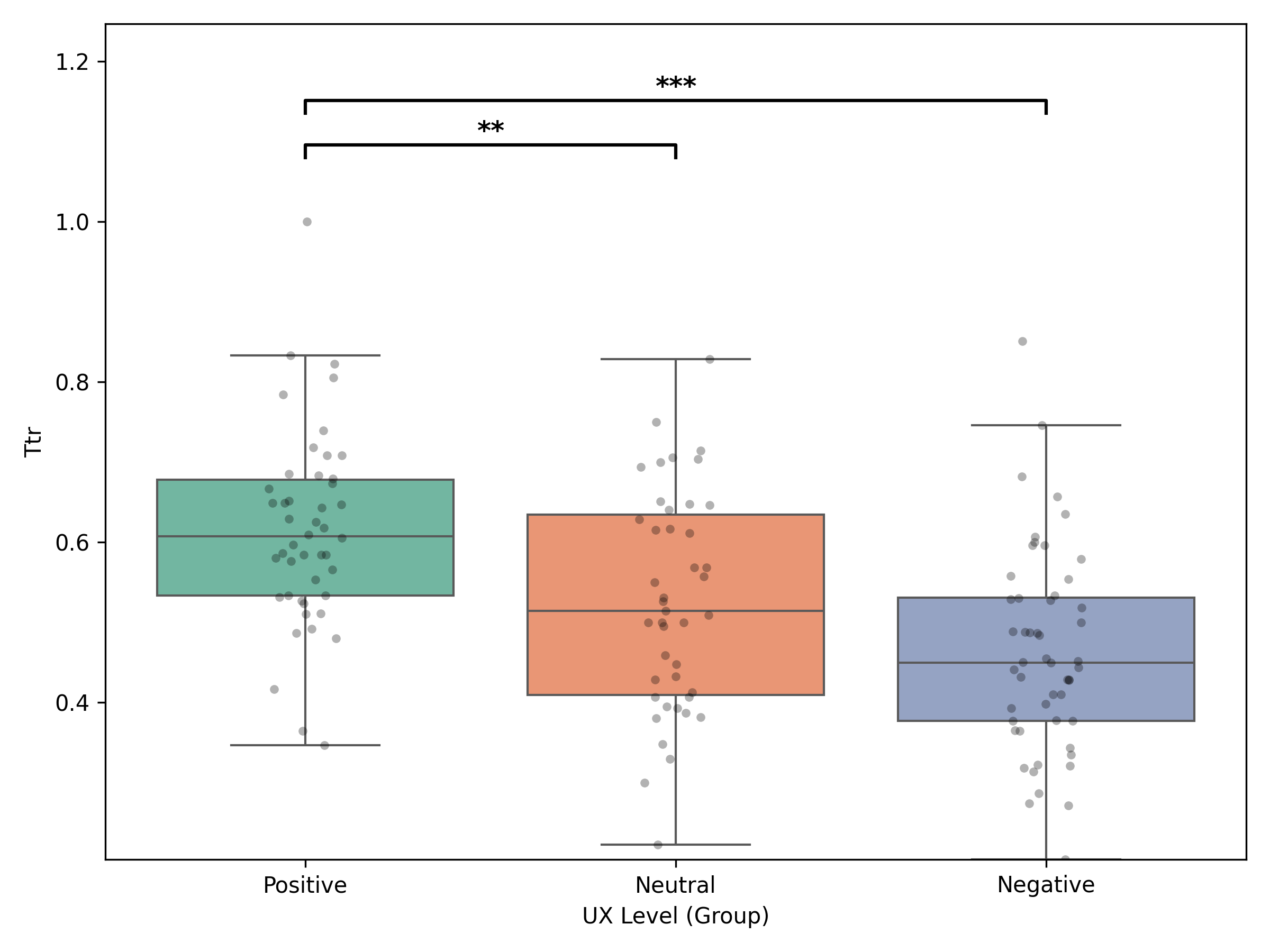}
        \caption{Lexical diversity (TTR)}
        \label{fig:ttr}
    \end{subfigure}

    \caption{Distributions of key lexical features across KPI-defined UX levels (Positive, Neutral, Negative). Higher type--token ratio and lexical diversity are associated with more positive UX, whereas lower scores and shorter average word length indicate more constrained or reduced language use during negative interactions. Significance markers reflect Kruskal--Wallis tests with Dunn--Bonferroni post-hoc comparisons.}
    \Description{Three side-by-side boxplots showing distributions of mean type--token ratio, average word length, and lexical diversity across positive, neutral, and negative user experience levels, with statistical significance annotations for group differences.}
    \label{fig:lexical_three_features}
\end{figure*}

To understand how linguistic expression varied with perceived interaction quality, we examined three lexical features, such as Type–Token Ratio (TTR), mean Type–Token Ratio, and average word length, across the three UX levels. A consistent pattern emerged: users experiencing more positive interactions produced more linguistically diverse utterances, whereas negative UX was associated with reduced lexical richness and shorter word choices. As shown in Figure~\ref{fig:lexical_three_features}, both TTR and mean Type–Token Ratio demonstrated highly significant group effects (TTR: $H=29.60$, $p < .001$, $\varepsilon^2=0.20$; mean TTR: $H=29.12$, $p < .001$, $\varepsilon^2=0.197$). Post-hoc comparisons revealed that Positive UX elicited substantially higher lexical diversity than both Neutral ($p = .0098$ for TTR; $p = .0118$ for mean TTR) and Negative UX ($p < .001$ for both metrics). Negative UX, in contrast, showed the lowest lexical complexity, suggesting shorter, more repetitive phrasing during frustrating or unsuccessful interactions. Average word length also differed significantly across UX groups ($H=7.46$, $p = .024$, $\varepsilon^2=0.04$). Dunn–Bonferroni tests indicated that users in the Positive UX condition produced longer words than those in the Negative UX group ($p = .034$), again pointing to reduced linguistic complexity under negative interaction states. Together, these results show that lexical behavior is tightly coupled with UX quality: positive interactions encourage more varied and expressive speech, while negative experiences correspond to simplified or compressed linguistic output, reflecting cognitive strain or disengagement during problematic voice-assistant interactions.



\section{Speech–UX Relationship Synthesis}
\label{speech_ux}
\subsection{Interpretability of Speech–UX Relationships}
To improve the interpretability of our findings, Table~\ref{tab:feature-ux-summary} provides a consolidated overview of how specific acoustic and linguistic speech features relate to user experience metrics. The table also reports effect sizes, clarifying the relative strength of each relationship. As shown, prosodic expressiveness (e.g., pitch and spectral variability) and timing behaviour (e.g., pause duration) exhibited the largest associations with UX. Voice quality perturbation measures showed the strongest effects, with loudness variability ($\epsilon^2 = 0.090$) and spectral flux variability ($\epsilon^2 = 0.093$) being particularly diagnostic of satisfaction. These findings support the idea that emotional intensity, fluency, and vocal control are meaningful indicators of a participant's moment-to-moment experience during voice-based interaction.

\begin{table*}[htbp]
\centering
\caption{Summary of Speech Feature Associations with UX Dimensions, Effect Sizes, and Statistical Significance}
\Description{Table summarizing significant associations between multiple speech feature categories and user experience dimensions, including effect sizes and p-values, indicating how specific acoustic, social, voice-quality, and paralinguistic features differ across UX levels.}
\label{tab:feature-ux-summary}
\begin{tabular}{@{}llllll@{}}
\toprule
\textbf{UX Dimension} & \textbf{Feature Category} & \textbf{Specific Feature} & \textbf{Association} & \textbf{Effect Size ($\epsilon^2$)} & \textbf{$p$-value} \\
\midrule
\textbf{Trust} & Time-Frequency & ZCR Mean & Negative > Positive & 0.040 & 0.0225 \\
 & Time-Frequency & Spectral Centroid Mean & Negative > Positive & 0.052 & 0.0102 \\
 & Time-Frequency & Spectral Bandwidth Mean & Negative > Positive & 0.064 & 0.0044 \\
 & Social Speech & Engagement Rate/min & Negative < Neutral & 0.040 & 0.0229 \\
 & Voice Quality & Loudness StdDev & Negative > Neutral & 0.049 & 0.0119 \\

\cmidrule(r){1-6}
\textbf{Attractiveness} & Time-Frequency & ZCR Mean & Negative > Positive & 0.039 & 0.0251 \\
 & Time-Frequency & Spectral Centroid Mean & Negative > Positive & 0.057 & 0.0071 \\
 & Social Speech & Engagement Rate/min & Negative < Positive & 0.034 & 0.0349 \\
 & Voice Quality & Spectral Flux StdDev & Negative > Positive & 0.031 & 0.0415 \\

\cmidrule(r){1-6}
\textbf{Satisfaction} & Time-Frequency & RMSE Mean & Negative < Positive & 0.032 & 0.0398 \\
 & Time-Frequency & Spectral Centroid Mean & Negative > Positive & 0.043 & 0.0190 \\
 & Social Speech & Duration (s) & Negative > Positive & 0.035 & 0.0313 \\
 & Social Speech & Activity Ratio & Negative < Positive & 0.058 & 0.0066 \\
 & Social Speech & Engagement Rate/min & Negative < Positive & 0.079 & 0.0015 \\
 & Voice Quality & Loudness StdDev & Negative > Positive & 0.090 & 0.0007 \\
 & Voice Quality & Spectral Flux StdDev & Negative > Positive & 0.093 & 0.0006 \\

\cmidrule(r){1-6}
\textbf{Comprehensibility} & Time-Frequency & Spectral Centroid Mean & Negative > Positive & 0.042 & 0.0193 \\
 & Social Speech & Activity Ratio & Positive > Neutral & 0.041 & 0.0215 \\
 & Voice Quality & Loudness StdDev & Negative > Positive & 0.067 & 0.0034 \\
 & Voice Quality & Spectral Flux StdDev & Negative > Positive & 0.088 & 0.0008 \\

\cmidrule(r){1-6}
\textbf{Response Behavior} & Social Speech & Activity Ratio & Neutral > Negative & 0.052 & 0.0099 \\
 & Social Speech & Engagement Rate/min & Neutral > Negative & 0.050 & 0.0116 \\
 & Voice Quality & HNRDBACF StdDev & Negative > Positive & 0.074 & 0.0021 \\

\cmidrule(r){1-6}
\textbf{Response Quality} & Social Speech & Duration (s) & Negative > Neutral & 0.090 & 0.0007 \\
 & Voice Quality & Loudness StdDev & Negative > Positive & 0.052 & 0.0102 \\
 & Voice Quality & Spectral Flux StdDev & Negative > Positive & 0.049 & 0.0120 \\
 & Voice Quality & HNRDBACF StdDev & Negative > Positive & 0.060 & 0.0058 \\
\bottomrule
\end{tabular}

\vspace{0.2cm}
\footnotesize 
\textit{Note:} Effect sizes ($\epsilon^2$) follow Cohen's conventions: .01 = small, .06 = medium, .14 = large. Association direction indicates which UX level shows higher feature values (e.g., "Negative > Positive" means Negative UX had significantly higher values than Positive UX). Only features with $p < .05$ after Dunn-Bonferroni correction are shown. ZCR = Zero-Crossing Rate; RMSE = Root Mean Square Energy; HNRDBACF = Harmonic-to-Noise Ratio; StdDev = Standard Deviation.
\end{table*}

\subsection{Individual Differences in Speech Patterns}
An important consideration for speech-based UX sensing is the extent to which individual differences in vocal style influence model performance. Speakers naturally differ in baseline pitch ranges, conversational pacing, prosodic expressiveness, and use of disfluencies. Our modeling approach partially mitigates this challenge by extracting relative within-speaker changes rather than absolute values, and through mixed-effects modeling that separates UX-related variance from stable speaker-specific traits. While our results demonstrate clear patterns across the current participant sample, future work with larger and more diverse populations is needed to assess how strongly idiosyncratic speech patterns constrain generalizability or whether adaptive or personalized models would further improve robustness.

\subsection{Gender Effects}
Although gender was not a primary factor in our study design, we conducted exploratory comparisons to evaluate whether gender systematically influenced either the extracted speech features or UX ratings. No reliable gender differences were observed in the patterns of feature change across UX levels. Participants of different genders exhibited comparable acoustic patterns and similar UX outcomes across the three personas. However, due to the modest and imbalanced sample, these analyses should be interpreted cautiously. Replication with larger and more balanced datasets would help clarify whether subtle gender-linked vocal characteristics interact with perceived UX in voice-based interaction contexts.

\section{Overview of Significant Acoustic, Social, Voice-Quality, and Paralinguistic Differences Across UX Levels}
\label{over_speech_table}

Table~\ref{tab:master_speech_table} summarizes all speech-related features that exhibited significant differences across the three UX levels (Positive, Neutral, Negative) based on Kruskal--Wallis tests with Dunn--Bonferroni post-hoc comparisons. Across feature categories, consistent patterns emerged linking degraded user experience to increased acoustic instability, reduced lexical richness, and altered interaction dynamics.
For time--frequency acoustic features, negative UX was associated with higher spectral dispersion, brightness, and noisiness, reflected by increases in spectral bandwidth, spectral centroid, and zero-crossing rate. These trends suggest more irregular and strained vocal production during poorer interaction experiences.
Social speech features revealed systematic behavioral shifts across UX levels. Participants exhibited higher speech activity and engagement rates during positive UX conditions, whereas negative UX was characterized by longer interaction durations, indicating potentially prolonged or effortful exchanges.
Voice-quality metrics derived from openSMILE further highlighted increased vocal instability during negative UX. Higher jitter, loudness variability, harmonic-to-noise ratio variability, pitch variability, and formant amplitude variability consistently distinguished negative from positive UX levels, suggesting greater vocal strain and reduced speech control during less satisfying interactions.
Finally, paralinguistic and lexical features showed strong effects of UX quality on language use. Positive UX was associated with higher lexical diversity and type--token ratios, whereas negative UX corresponded to shorter average word lengths and reduced linguistic richness, indicating more constrained verbal expression during unfavorable interactions.

Overall, the results demonstrate that UX quality is reflected not only in subjective ratings but also in measurable changes across acoustic, behavioral, and linguistic speech characteristics, providing converging multimodal evidence for systematic UX-related speech patterns.

\begin{table*}[ht]
\centering
\caption{Summary of statistical analysis across all acoustic, social, voice-quality, and paralinguistic speech features. Only significant Kruskal--Wallis tests are shown, along with Dunn–Bonferroni pairwise comparisons indicating direction of UX‐related effects.}
\Description{Comprehensive table summarizing significant Kruskal--Wallis test results across multiple speech feature categories, including effect sizes and post-hoc Dunn--Bonferroni comparisons highlighting differences between positive, neutral, and negative user experience levels.}
\label{tab:master_speech_table}
\small
\begin{tabular}{llcccc}
\toprule
\textbf{Category} & \textbf{Feature} & \textbf{Kruskal–Wallis (H)} & 
\textbf{$p$} & \textbf{Effect Size ($\varepsilon^2$)} & 
\textbf{Significant Dunn--Bonferroni Comparisons} \\
\midrule

\multicolumn{6}{l}{\textbf{Time--Frequency Acoustic Features}} \\
\midrule
TF Acoustic & ZCR Mean & 9.62 & .008 & .055 & Pos $<$ Neg ($p=.017$) \\
TF Acoustic & Spectral Centroid Mean & 13.71 & .001 & .084 & Pos $<$ Neu ($p=.005$), Pos $<$ Neg ($p=.003$) \\
TF Acoustic & Spectral Bandwidth Mean & 11.51 & .003 & .068 & Pos $<$ Neu ($p=.0047$), Pos $<$ Neg ($p=.0226$) \\

\midrule
\multicolumn{6}{l}{\textbf{Social Speech Features}} \\
\midrule
Social & Activity Ratio & 7.141 & .028 & .037 & Pos $>$ Neg ($p=.0262$) \\
Social & Duration (s) & 6.667 & .036 & .034 & Neg $>$ Pos ($p=.0302$) \\
Social & Engagement Rate/min & 7.328 & .026 & .038 & Pos $>$ Neg ($p=.0569$) \\

\midrule
\multicolumn{6}{l}{\textbf{Voice Quality Metrics (openSMILE eGeMAPS)}} \\
\midrule
Voice Quality & Intensity Mean & 10.712 & .0047 & .063 & Pos $>$ Neg ($p=.007$), Pos $>$ Neu ($p=.0293$) \\
Voice Quality & F0 Variability (semitone) & 10.413 & .0055 & .061 & Pos $<$ Neu ($p=.0145$), Pos $<$ Neg ($p=.0159$) \\
Voice Quality & Loudness Variability & 9.362 & .0093 & .053 & Pos $<$ Neg ($p=.012$) \\
Voice Quality & Jitter Local Variability & 9.791 & .0075 & .056 & Neu $<$ Neg ($p=.0061$) \\
Voice Quality & HNR Variability & 9.666 & .0080 & .055 & Pos $<$ Neg ($p=.0061$) \\
Voice Quality & F1 Amplitude (log‐rel‐F0) & 12.197 & .0022 & .073 & Pos $<$ Neu ($p=.0295$), Pos $<$ Neg ($p=.0025$) \\

\midrule
\multicolumn{6}{l}{\textbf{Paralinguistic / Lexical Features}} \\
\midrule
Paralinguistic & Avg. Word Length & 7.459 & .024 & .040 & Neg $<$ Pos ($p=.0343$) \\
Paralinguistic & TTR (Lexical Diversity) & 29.124 & $<.001$ & .197 & Pos $>$ Neu ($p=.0118$), Pos $>$ Neg ($p<.001$) \\
Paralinguistic & Mean Type–Token Ratio & 29.599 & $<.001$ & .200 & Pos $>$ Neu ($p=.0098$), Pos $>$ Neg ($p<.001$) \\

\bottomrule
\end{tabular}
\end{table*}

\end{document}